\documentclass[%
reprint,
amsmath,amssymb,
aps,
prb
]{revtex4-2}

\usepackage{graphicx}
\usepackage{dcolumn}
\usepackage{bm,bbm}

\newcommand {\Svec}{{\mathbf{S}}}
\newcommand {\Hvec}{{\mathbf{H}}}
\newcommand{\Heff}{{\mathbf{H}}_\text{eff}}

\newcommand {\Mvec}{{\mathbf{M}}}
\newcommand {\Vvec}{{\mathbf{V}}}
\newcommand {\Fvec}{{\mathbf{F}}}

\newcommand {\Wvec}{{\mathbf{W}}}
\newcommand {\Gvec}{{\mathbf{G}}}

\newcommand {\xvec}{{\mathbf{x}}}
\newcommand {\evec}{{\mathbf{e}}}
\newcommand {\rvec}{{\mathbf{r}}}
\newcommand {\mvec}{{\mathbf{m}}}
\newcommand {\svec}{{\mathbf{s}}}
\newcommand {\vvec}{{\mathbf{v}}}

\newcommand {\kvec}{{\mathbf{k}}}

\newcommand {\hvec}{{\mathbf{h}}}

\newcommand {\varphivec} {\boldsymbol{\varphi}}

\newcommand {\nvec}{{\mathbf{n}}}

\newcommand {\Dvec}{{\bm{D}}}
\newcommand {\Hvecm}{{\mathbf{H}_\text{m}}}
\newcommand {\Hveca}{{\mathbf{H}_\text{a}}}

\begin{document}
	
	\preprint{APS/123-QED}
	
	\title{Quantum Micromagnetic Theory of Magnons in Finite Nanostructures } 
	
	\author{Claudio Serpico}
	\author{Salvatore Perna}%
	\author{Massimiliano d'Aquino}
	\affiliation{
		Department of Electrical Engineering and Information Technologies
		University of Naples Federico II, Italy
	}
	
	
	\date{\today}
	
\begin{abstract}
This paper presents a quantum field theoretical formalism for studying magnons in finite nanostructures with arbitrary shapes and spatially nonuniform ground states. It extends the classical micromagnetic formalism by introducing a micromagnetic Hamiltonian quantum operator, which incorporates exchange, Dzyaloshinsky-Moriya, anisotropy, magnetostatic, and Zeeman energies. The nonuniformity of the ground state is handled by pointwise aligning  the quantization axis of the magnetization field operator  with the classical ground state. The Hamiltonian is expanded in the large spin-number limit and truncated to retain only terms quadratic in the components of the magnetization operator transverse to the
quantization axis. This quadratic Hamiltonian is used to derive the linear quantum Landau-Lifshitz equation. 
By diagonalizing this equation under appropriate boundary and normalization conditions, a discrete set of magnon creation and annihilation operators is obtained, 
enabling a complete description of the magnon spectrum. 
Finally, the theory is applied to study  the effects of temperature and shape on low-temperature thermal equilibrium fluctuations of magnons in thin ferromagnetic nanodisks.
\end{abstract}
	
	\maketitle

\section{Introduction}
Magnons are quasiparticles associated to quantized spin waves,  that are collective excitations of coupled spins in magnetic solids
\cite{Bloch_1930, Bloch_1932, Holstein_Primakoff_1940, Herring_Kittel_1951, Dyson_1956, Kittel_book_1963, Akhiezer_book, Stancil_Prabhakar_book, Sparks_book}.  
They play a crucial role in the study of ferromagnetic materials influencing the thermodynamic 
properties of the media and controlling the dynamics of magnetization as a response to various external 
stimuli, such as microwave fields\cite{ Stancil_Prabhakar_book}, spin-transfer-torque\cite{Urazhdin_2014} and spin-orbit effects\cite{Demidov_2020}. The study of magnons generation, manipulation, and control has given rise to the 
field of magnonics\cite{Lenk_2011, Barman_2021},  which focuses on the use of spin waves in magnetic 
nanostructures. 
Magnonics intersects with other key nanotechnological research areas, such as spintronics\cite{Chumak_2015}, nanophotonics\cite{Chernov_2020}, 
and plasmonics\cite{Maksymov_2016}. In recent decades, the growing interest in quantum science and technology has also led to 
the use of magnons as carriers of quantum information, resulting in the emerging field of quantum magnonics\cite{Yuan2022}.

The quantum theory of spin waves has a long history, originating from the seminal  work of Bloch, Holstein, Primakoff,  Kittel 
and Dyson\cite{Bloch_1930, Bloch_1932, Holstein_Primakoff_1940, Herring_Kittel_1951, Dyson_1956, Kittel_book_1963}.
The formalism they developed, based on the concept of  plane waves, remains widely used today. While this approach 
works well for bulk crystals with spatially uniform ground states and negligible boundary effects, it is less 
suitable for finite magnetic nanostructures. In finite systems, even when material properties are spatially homogeneous, 
spatially nonuniform (noncollinear) ground states often arise due to demagnetizing fields at the boundaries. Additionally, 
spatial nonuniformities can result from Dzyaloshinsky-Moriya interactions\cite{Dzyaloshinskii1958, Moriya1960, Bogdanov1994}, 
which may lead to the formation of magnetic Skyrmions\cite{Fert2017}.
	
In this paper, we introduce a formalism in which the role of plane waves is replaced by the magnetization normal modes. 
The proposed approach consists in the quantum extension of classical micromagnetics, that is the continuum theory 
for ferromagnetism \cite{Brown_1963}. It is based on a general quantum Hamiltonian that incorporates all relevant interactions, 
including exchange, Dzyaloshinsky-Moriya (DM), dipolar,  anisotropy, and Zeeman effects. 
The formalism enables the self-consistent derivation of the magnon spectrum associated with
magnetization normal modes, while fully accounting for boundary conditions and noncollinear ground states.
	
The classical (non-quantum) treatment of normal modes in finite structures has traditionally relied on micromagnetic theory. 
A seminal contribution in this area came from Walker\cite{Walker_1957}, who analytically derived magnetostatic normal modes for spheroids. Walker's approach was later extended to ellipsoidal geometries and further refined to include exchange interactions\cite{Aharoni_1963, Arias_2001, Arias_2004, Arias_2005a, Arias_2005b}. However, this analytical treatment was limited to cases involving spatially uniform ground states (saturated samples).
On the other hand, extensive classical studies of dipole-exchange normal modes in ferromagnets with 
noncollinear ground states and arbitrary shapes have been conducted using numerical methods, including finite difference and finite element 
techniques \cite{Labbe_1999, Vukadinovic_200, Grimsditch_2004a, Grimsditch_2004b, Rivkin_2004, dAquino_2009, McMicheal_2005}. 
Notably, 
the numerical methods presented in Ref.\cite{dAquino_2009} are developed 
from a general operator-based formulation of the continuum classical normal modes problem, 
which is connected to the quantum treatment in the present paper. 
	
General aspects of the quantum extension of classical micromagnetic theory were explored by Mills\cite{Mills_2006},\cite{Mills_2007}. His formalism incorporates the finite size of the system through appropriate boundary conditions but is based on the simplifying assumption of spatially uniform ground states.
On the other hand, noncollinear ground states in magnetic systems have been extensively studied within the context of Heisenberg-type spin models. Foundational work in this area dates back to the 1950s, particularly in relation to antiferromagnets, spiral, and helical magnets (see Ref. \cite{Rastelli_2013} and references therein).
Over the past two decades, interest in materials exhibiting noncollinear magnetic ground states has resurged, 
in the field of magnetic nanotechnologies \cite{Diep_2022}. Recent investigations in this area have examined 
Heisenberg and Haldane-gap spin rings \cite{Schuetz_2003, Schuetz_2004a, Schuetz_2004b}, quantum antiferromagnets \cite{Spremo_2005, Kreisel_2008}, 
quantum spin waves in finite chiral spin chains \cite{Roldan-Molina_2014}, quantum fluctuations around skyrmionic ground states \cite{Roldan-Molina_2015}, 
and spin currents and spin superfluidity \cite{Rueckriegel_2017}.
These studies often rely on the assumption, used also in the present paper, that the system operates under conditions where 
quantum and thermal fluctuations are sufficiently small, justifying the treatment of low-energy quantum excitations 
as occurring near the classical ground state. Based on this assumption, the quantization procedure uses the direction of the classical noncollinear 
ground state as the quantization axis for the spin (or magnetization) quantum operators\cite{Rastelli_2013}, \cite{Haraldsen_2009}.

The quantum extension of the micromagnetic formalism presented in this paper is structured around two primary 
novel aspects. 
First, we introduce a formulation for analyzing magnon spectrum using the most general form 
of the Hamiltonian, rather than tailoring it to specific material properties or geometries, as is commonly 
done in the studies on the subject \cite{Rastelli_2013}-\cite{Rueckriegel_2017}. The Hamiltonian operator 
considered, essentially the quantum version of the micromagnetic energy functional 
\cite{Brown_1963, Aharoni_Ferromagnetism}, incorporates all relevant interactions among spins within a magnetic body of arbitrary shape, explicitly accounting for the often-overlooked 
nonlocal contributions of magnetostatic fields. Notably, these fields are recognized as a significant source 
of computational complexity when addressing the diagonalization problem.
Second, the diagonalization procedure is based on the quantum version of the Landau-Lifshitz equation 
linearized around the ground state \cite{Stancil_Prabhakar_book}, \cite{Brown_1963}. This approach, while equivalent to the conventional method based on the
combination of Holstein-Primakoff transformation and Bogoliubov diagonalization \cite{Holstein_Primakoff_1940}, \cite{Sparks_book}, has the advantage of maintaining 
boundary conditions  in their simplest form. The resulting diagonalization procedure gives rise to a generalized eigenvalue problem similar to 
those traditionally encountered in classical micromagnetic analyses \cite{dAquino_2009}. This connection allows for the application of 
well-established micromagnetic computational techniques \cite{Abert_2019} and provides guidance 
on how to extend numerical codes developed for the classical case to the quantum scenario. 
In this context, an important aspect that arises is the normalization conditions that must be imposed on 
eigenfunctions describing the spatial profile of quantum magnons to preserve canonical commutation relations. 
The proposed formalism leads to an extension of Walker-type normalization conditions, as discussed in 
Refs. \cite{Walker_1957, Mills_2006}, to the more general case where ground state is spatially nonuniform,
and, in addition, DM interactions and anisotropic
exchange are taken into account.

The effectiveness of the proposed theory is demonstrated in the final part of the paper, where numerical calculations of thermal equilibrium averages associated with the Cartesian components of magnetization in a thin disk are presented.
For an in-plane magnetized film, the computations reveal that at low temperatures, edge magnons produce thermal fluctuations near the material boundaries that are significantly larger than those predicted by the plane-wave analysis for an infinitely in-plane magnetized thin film. In the case of a thin disk with a Skyrmion ground state, this approach enables the analysis of the spatially nonuniform distribution of thermal fluctuations, showing that these fluctuations are particularly enhanced in regions with the steepest gradients in the ground-state magnetization.
These calculations, which can be applied to bodies of arbitrary shapes and noncollinear ground states, demonstrate the formalism's ability to study thermal equilibrium magnon populations. This capability is crucial for engineering magnetic nanostructures for low-temperature quantum magnonics applications in quantum computing and information processing \cite{Yuan2022}, \cite{Yuan2022a}.

The paper is structured as follows. 
Sections \ref{sec:spin_hamiltonian}, \ref{sec:Classical_Ground_State},
\ref{eq:expansion_of_Hamiltonian}, \ref{sec:Quantum Micromagnetic Spin-waves} present the quantum micromagnetic formalism 
for finite nanostructures with arbitrary shapes and spatially nonuniform ground states, culminating in the derivation of 
quantum linear Landau-Lifshitz equations. 
Section \ref{sec:Diagonalization of Quantum Magnetization Dynamics}
is dedicated to the procedure for determining the magnon spectrum, establishing the conditions for diagonalizing the equation 
of motion, and introducing creation-annihilation operators to describe magnetization dynamics via magnonic quasi-particles.
In this section Walker-type normalization conditions required for diagonalization are also derived.
 In Section \ref{sec:applications}
the formalism is used to calculate low-temperature thermal equilibrium fluctuations of magnons in nanoscale thin disks. Finally, 
Section \ref{sec:conclusions} presents conclusions  and discusses possible future extensions of the work.

\section{Quantum Micromagnetic Hamiltonian}
	\label{sec:spin_hamiltonian}

We begin by considering a ferromagnetic Heisenberg-type model 
and proceed by taking its continuum limit, following the approach 
outlined in Refs.\,\cite{Kittel_book_1963},
\cite{Akhiezer_book}.
In the Heisenberg model, the ferromagnet
is described by the interaction of $N$ spin-$S$ particles 
arranged into a finite  three-dimensional lattice.
The spin angular momentum operator (measured in units of $\hbar$) 
of the particle in the position $\rvec_i$ in the lattice  is
\begin{align}
	\label{eq:vector_spins}
	\hat{\Svec}_i = \hat{S}^1_i \evec_1 + \hat{S}^2_i \evec_2+ \hat{S}^3_i \evec_3, 
\end{align}
where $(\evec_1,\evec_2,\evec_3)$ are Cartesian unit vectors and $i=1,\ldots,N$.  
The operators $\hat{S}_i^u$ satisfy the 
	conditions
	\begin{align}
		\label{eq:Spin_commutation_relations}
		&
        [\hat{S}^u_i,\hat{S}_i^v] = i \epsilon_{uvr} \hat{S}^r_i\, ,  \\
		\label{eq:S^2-relation}
		&\hat{\Svec}_i^2 = \hat{\Svec}_i\cdot \hat{\Svec}_i= S(S+1)\hat{\mathbbm{1}}  \, , \\
        \label{eq:Spin_commutation_relations_i_j}
        &  [\hat{\Svec}_i,\hat{\Svec}_j] = 0 \quad  \text{with} \, i\neq j \, ,
	\end{align}
	where $u,v,r=1,2,3$,  $[\,\cdot\,, \cdot\, ]$ is the quantum commutator, $i,j=1, \ldots N$, $\epsilon_{uvr}$ is the Levi-Civita 
	symbol, $\delta_{ij}$ the Kronecker delta and $\hat{\mathbbm{1}}$ is the identity operator  in the Hilbert state space 	associated to the spin system. In equation \eqref{eq:Spin_commutation_relations}, and throughout this paper, we adopt the Einstein summation convention for repeated indices.
	The transition to the continuum theory is based on the introduction of the field $\hat{\svec}(\xvec)$ representing the density of 	
	spin-angular momentum per unit volume that is connected to the Heisenberg Spins by the relation: 
	\begin{align}
		\label{eq:connections_s_and_S}
		\int_{V_c(\rvec_i)} \hat{\svec}(\xvec^\prime) \, dV_{\xvec^\prime} = \hat{\Svec}_i 	\, ,
	\end{align}
	where $V_c(\rvec_i)$ is the primitive cell of the lattice centered at $\rvec_i$.
	Then, the magnetization quantum field operator 
	$\hat{\Mvec}(\xvec)$ is given by the following formula
	\begin{align}
		\label{eq:def_of_Quantum_M}
		\hat{\Mvec}(\xvec) = -|g|\mu_B \hat{\svec}(\xvec) \, .
	\end{align}
    where $g \approx -2$ the Land\'e factor, 
    $\mu_B = |e| \hbar/(2 m_e) \approx 9.2740 \cdot 10^{-24} \, \text{Am}^2$ is the Bohr magneton ($e,m_e$ are the
    charge and the mass of the electron, respectively).
   	The maximum value of magnetization of the ferromagnet is
    \begin{align}
    	\label{eq:maximum_magnetization_M0}
    	M_0 = |g| \mu_B S/|V_c| \, ,
    \end{align}
   where   $|V_c|$ (typically in the order of
   $10^{-27} \div 10^{-28}$ m$^{-3}$ ) is the
   volume of the primitive cell of the lattice.
    Equations \eqref{eq:def_of_Quantum_M}, \eqref{eq:maximum_magnetization_M0} can be alternatively written
    as 	$\hat{\Mvec}(\xvec) = -\gamma \hbar \hat{\svec}(\xvec)$ 
    and $M_0 = \gamma \hbar S /|V_c|$, where  $\hbar$ is the Planck's constant and 
    $\gamma = |{g}| \mu_B /\hbar \approx (2\pi) \cdot 28.03 \cdot 10^9\,\text{rad T}^{-1} $
    is the gyromagnetic ratio.  Typical values of $\mu_0 M_0$ ($\mu_0=4\pi \cdot 10^{-7}$ H/m is the vacuum magnetic permeability)
    are in the order of one Tesla.
    By using eqs. \eqref{eq:Spin_commutation_relations}-\eqref{eq:def_of_Quantum_M}, we can infer
    the following relations for the magnetization:
	\begin{align}
		\label{eq:commu_rels_on_M_a}
		&	[\hat{M}^u(\xvec),\hat{M}^v(\xvec')] = -i \epsilon_{uvr}{\gamma \hbar}\hat{M}^r(\xvec)\delta(\xvec-\xvec^\prime) \, ,\\
		\label{eq:commu_rels_on_M_b}
		&	(\hat{\Mvec}(\xvec))^2  
		= M_0^2 \left(1+1/S \right) \hat{\mathbbm{1}}\, ,
	\end{align}
that are the continuum counterpart of eqs.\eqref{eq:Spin_commutation_relations}-\eqref{eq:Spin_commutation_relations_i_j}.

The central quantity in the quantum micromagnetic theory is  the following Hamiltonian operator that can be obtained as the continuum limit of the ferromagnetic 
Heisenberg Hamiltonian\cite{Kittel_book_1963},\cite{Akhiezer_book}:
	\begin{widetext}
		\begin{align}
			\label{eq:magnetization_hamiltonian}
			\hat{H}= 
			\int_\Omega  dV  \left\{
			\frac{A_{kh}^{uv}}{M_0^2}\, \frac{\partial \hat{M}^u}{\partial x_k}\frac{\partial \hat{M}^v}{\partial x_h}  
			+\frac{1}{2}{\frac{ D_{k}^{uv}}{M_0^2}} \left( \! \hat{ M}^u \frac{\partial \hat{M}^v}{\partial x_k}- \frac{\partial \hat{M}^u}{\partial x_k} \hat{ M}^v  \!\right) 
			- \hat{M}^u\frac{K^{uv}}{M_0^2} \, \hat{M}^v  
            - \frac{\mu_0}{2}\hat{\Mvec} \cdot \Hvecm[\hat{\Mvec}] -\mu_0 \hat{\Mvec} \cdot \Hveca  \right\} ,
		\end{align}
	\end{widetext}
    where 
    superscripts $u,v\ = 1,2,3$ are used to denote
    the Cartesian components of $\hat{\Mvec}$ 
    while $k,h=1,2,3$  are used to denote Cartesian components of $\xvec$.
    The upper and lower position of the indices does not have a special meaning in
    this context and the `$\cdot$' symbol denotes the usual Euclidean scalar product.
    
    The sum of the first two terms in the integrand at the right-hand-side of eq.\eqref{eq:magnetization_hamiltonian} 
    is the {exchange energy} density and its expression can be obtained by the appropriate
    continuum (long wavelength) limit of the Heisenberg Hamiltonian\cite{Kittel_book_1963}. The tensorial nature 
    of the parameters $A_{kh}^{uv}$  and $D_{k}^{uv}$ is considered to take into account 
    interplay between exchange and spin-orbit coupling \cite{Moriya1960}. In this respect,
    the tensor  $A_{kh}^{uv}$ is separately symmetric 
    with respect to exchange of indices $u,v$ and $k,h$, while 	$D_{k}^{uv}$ is anti-symmetric with respect to indices $u,v$.
   The second component of the exchange term represents the Dzyaloshinsky-Moriya exchange energy density \cite{Dzyaloshinskii1958}, \cite{Moriya1960}, expressed in a symmetrized form to ensure that the energy density remains Hermitian. 
   In typical micromagnetic studies, isotropic exchange is considered, where $A^{uv}_{kh} = A \delta_{uv}\delta_{kh}$ and $A$ denotes 
   the exchange constant. 
   The third term in  eq.\eqref{eq:magnetization_hamiltonian} 
    is the anisotropy energy density, characterized by the symmetric tensor $K^{uv}$. In cases of uniaxial anisotropy, often 
    assumed in micromagnetics, $K^{uv}$ takes the dyadic form $K^{uv} = K_1 e^u_\text{an}e^v_\text{an}$, where $K_1$
    is the anisotropy constant, and where the unit vector $\evec_\text{an}$ aligns with the anisotropy axis. 
     The quantities $A_{kh}^{uv}$, $D_{k}^{uv}$, and $K^{uv}$, characterizing the
	material properties of the body, are possibly depending on the position. This dependence
	will be understood throughout the paper.
	
	In the last two terms in the integrand  at the right-hand-side of eq.\eqref{eq:magnetization_hamiltonian}, 
	$\Hvecm[\hat{\Mvec}]$ denotes the magnetostatic 
	field operator that is 	defined as
\begin{equation}
		\label{eq:magnetostatic_field}
		\Hvecm[\hat{\Mvec}](\xvec) =  \evec_u \int_{\Omega} N^{uv}(\xvec-\xvec^\prime) \hat{M}^v(\xvec^\prime) \, dV_{\xvec^\prime} \, ,
\end{equation}
where
\begin{align}
	\label{eq:Green_fun_Dyadics_0}
	N^{uv}(\xvec) =  \delta_{uk}\delta_{vh} \frac{\partial}{\partial x_k} \frac{\partial}{\partial x_h} \left( \frac{1}{4\pi |\xvec|} \right)\, ,
\end{align}
and $\Hveca$ is the applied magnetic field. The corresponding terms in the Hamiltonian are the 
	magnetostatic energy and the Zeeman energy, respectively. The magnetostatic energy has an Hermitian
	density as a consequence of the symmetry of dyadic kernel \eqref{eq:Green_fun_Dyadics_0} with respect to the indices 
	$u$ and $v$ and the exchange between $\xvec$ and $-\xvec$.

	Quantum magnetization dynamics in the Heisenberg picture associated to the Hamiltonian operator 
	\eqref{eq:magnetization_hamiltonian} is governed
	by the following equation:
	\begin{align}
		\label{eq:Heisenberg_M}
		i\hbar \frac{\partial} {\partial t} \hat{M}^u(\xvec,t)  = \left[\hat{M}^u(\xvec,t) \, , \,   
		\hat{H} \right] \, ,
	\end{align}
	where  $\hat{M}^u(\xvec,t)$, with $u=1,2,3$ are operators 
	in the Heisenberg picture. The   operators $\hat{M}^u(\xvec)$ in the Schroedinger picture are related to $\hat{M}^u(\xvec,t)$ 
	by the initial condition  $\hat{M}^u(\xvec)=\hat{M}^u(\xvec,t=0)$.
	
The quantum micromagnetic field model has been introduced as the continuum limit of a 
Heisenberg-type model, and thus it is expected to be applicable to insulating materials like Yttrium 
Iron Garnet (YIG), where magnetic moments are localized. In contrast, for metals such 
as iron (Fe), cobalt (Co), and nickel (Ni), the dynamics of conduction electrons significantly 
influence magnetization, a contribution that the Heisenberg model does not fully account for. 
However, the micromagnetic field theory can also be introduced independently of a specific spin 
model by relying on symmetry considerations. This partly justifies the phenomenological 
application  of the  theory to materials where itinerant electrons influence 
magnetization processes \cite{Kittel_book_1963}.

\section{Classical Ground state}
\label{sec:Classical_Ground_State}

We want to study physical situations where $S$ is sufficiently large,  
and the temperature $T$  is significantly lower than the Curie temperature. 
In  these conditions, both thermal and quantum fluctuations can be considered small 
and quantum states of the system are those  for which $\hat{\Mvec}(\xvec)$ has nearly 
maximal projection along the classical ground state. This state can be determined
by the classical micromagnetic theory,  where magnetization field operator $\hat{\Mvec}(\xvec)$ 
is treated as a classical field $\Mvec(\xvec)$ with a  given magnitude $|\Mvec(\xvec)| = M_0$.
The classical ground state $\Mvec_0(\xvec)$ is determined by minimizing the Hamiltonian
\eqref{eq:magnetization_hamiltonian} computed as function of  $\Mvec(\xvec)$, denoted by $H[\Mvec]$.
The classical fields that are minima of ${H}[\Mvec]$ are such 
that the first variation of  ${H}[\Mvec]$ vanishes
under the constraint $|\Mvec(\xvec)| = M_0$.
This leads to Brown's equations for micromagnetic equilibrium states \cite{Brown_1963}:
	\begin{align}
		\label{eq:equilibium_cond_1_m}
		\Mvec_0(\xvec) \times \Heff[\Mvec_0](\xvec) = 0  \, , \qquad \text{in} \,\,\, \Omega \, ,  
	\end{align}
\begin{align}
		\label{eq:bc_m_eq}
		n_k\left(  {2 A_{kh}^{uv}} \frac{\partial\,\, }{\partial x_h} +  { D_{k}^{uv}} \right){M}_0^u(\xvec)  = 0  \qquad \text{on} \,\,\, \partial \Omega  \, ,
\end{align}
with $v=1,2,3$, and	where $n_k$ are the cartesian components of  $\nvec$, the outward-pointing unit normal  to $\partial \Omega$
	(that is the surface bounding $\Omega$).
	Equation \eqref{eq:bc_m_eq} expresses the exchange energy natural  boundary conditions.
	The effective field $\Heff[\Mvec]$  in eq.\eqref{eq:equilibium_cond_1_m} is defined as
	\begin{align}
		\nonumber
		&\Heff[{\Mvec}] = -\frac{1}{\mu_0}\frac{\delta {H}}{\delta {\Mvec} } = \\ 
		\label{eq:Heff_0_1_a}
		&\qquad =\Hvec_\text{ex}[{\Mvec}] + \Hvec_\text{an}({\Mvec})+ \Hvecm[{\Mvec}] + \Hveca \, ,
	\end{align}
	where $\delta { H}/\delta {\Mvec}$ denotes the functional derivative of ${H}[\Mvec]$ with respect to ${\Mvec}$, and where
	\begin{align}
	\nonumber
		&\Hvec_\text{ex}[ {\Mvec}] =
		\evec_v\left[ \frac{\partial}{\partial x_k}\left(
		{ \frac{2 A_{kh}^{uv}}{\mu_0 M_0^2}} \frac{\partial {M}^u}{\partial x_h} \right)  
		+ \right.\\
			\label{eq:exchange_field}
	&	\left.
		\qquad \qquad\quad \frac{\partial}{\partial x_k}\left(  \frac{D_{k}^{uv}}{\mu_0 M_0^2} M^u \right)
		+ \frac{D_{k}^{uv}}{\mu_0 M_0^2} \frac{\partial M^u }{\partial x_k } \right]  \, ,\\
		\label{eq:anisotropy_field}
		& \Hvec_\text{an}({\Mvec}) = 
		\evec_v \left(  \frac{2K^{uv}}{\mu_0 M_0^2}   {{M}^u}\right) \, ,
	\end{align}
	are the {exchange} and {anisotropy} fields, respectively.
	
    The equilibrium equation \eqref{eq:equilibium_cond_1_m} is equivalent 
    to the condition
    \begin{equation}\label{eq:Brown_equilibrium}
    	\Heff[\Mvec_0](\xvec) =
    	\lambda_0(\xvec) \Mvec_0(\xvec) \, ,
    \end{equation}
    where  $\lambda_0(\xvec) = \Mvec_0(\xvec) \cdot \Heff[\Mvec_0]/M_0^2$.
     Examples of typical ground state configurations in thin nano-disks 
     can be seen in Figs.\,\ref{fig:Disk_geometry_and_groud_state_uniform},
    \ref{fig:Disk_geometry_and_groud_state_Skyrmion}.
    
	Notice that, on the basis of the expressions above, we can write the
	effective field in the form
	\begin{align}
		\label{eq:effective_field_op_form}
		\Heff[{\Mvec}](\xvec) = -\Dvec[{\Mvec} ](\xvec) + \Hveca(\xvec) 
	\end{align}
	where $-\Dvec[{\Mvec} ]$ is the linear operator 
	corresponding the first three terms at the right-had-side
	of eq.\eqref{eq:Heff_0_1_a} (see eq.\eqref{eq:Dvec_operator}).

	\section{Expansion of the Hamiltonian for large $S$}
	\label{eq:expansion_of_Hamiltonian}

	Now, we want to consider the expansion of the Hamiltonian \eqref{eq:magnetization_hamiltonian}
	for large values of $S$. The first step in our derivation is to take 
	the direction of the classical ground state, described by the unit vector
	\begin{align}
		\mvec_{0}(\xvec)  = \Mvec_{0}(\xvec)/M_0 \, ,
	\end{align}
	as quantization axis for $\hat{\Mvec}(\xvec)$\cite{Rastelli_2013}. We also introduce the 
	unit vectors $(\evec_{01}(\xvec), \evec_{02}(\xvec))$ orthogonal 
	to $\mvec_{0}(\xvec)$  that satisfy the conditions
	\begin{align}
		\label{eq:unit_vectors_orientations_m1}
		&\evec_{01}(\xvec) \cdot \evec_{02}(\xvec) = 0 \, , \\
		\label{eq:unit_vectors_orientations_m2}
		&\evec_{03}(\xvec)=\evec_{01}(\xvec) \times \evec_{02}(\xvec) = \mvec_{0}(\xvec) \, ,
	\end{align}
	such that the three vectors $(\evec_{01}(\xvec),\evec_{02}(\xvec), \evec_{03}(\xvec))$ form a
	system of Cartesian unit vectors.
	
The components of the magnetization operator $\hat{\Mvec}(\xvec)$ along the unit vectors $\evec_{0a}(\xvec)$,
denoted by $\hat{M}^{0a}(\xvec)$ for $a=1,2,3$, are related to the
components in the original Cartesian frame by the equation
\begin{align}
	\label{eq:Local_rotation_matrix}
	\hat{M}^u(\xvec) = e^{u}_{0a} (\xvec) \hat{M}^{0a}(\xvec)
\end{align}
where  $u,a=1,2,3$,  and $e^{u}_{0a}(\xvec)= \evec_u \cdot \evec_{0a}(\xvec)$  is the orthogonal matrix 
associated to the change of cartesian frame.

The three operators $\hat{M}^{0a}(\xvec)$, with $a=1,2,3$,
satisfy the same relations satisfied by the components
$\hat{M}^u(\xvec)$ expressed in eqs.\eqref{eq:commu_rels_on_M_a}, \eqref{eq:commu_rels_on_M_b}.

In the following, we make use of the decomposition
	\begin{align}
		\label{eq:Mvec_i_decomp}
		\hat{\Mvec}(\xvec) =  \hat{M}^{03}(\xvec) \, \mvec_{0}(\xvec) +  \hat{\Mvec}_{\bot}(\xvec) \, , 
	\end{align}
where
	\begin{align}
		\label{eq:Mbot_eo1_eo2}
		\hat{\Mvec}_{\bot}(\xvec) =  \hat{M}^{01}(\xvec) \evec_{01}(\xvec)+ \hat{M}^{02}(\xvec) \evec_{02}(\xvec) \, ,
	\end{align}
is the magnetization quantum field in the direction transversal to the quantization axis.

In the limit of large $S$ and temperatures much below Curie's temperature,
the quantum states of interest are those for which the quantum operator
$\hat{\Mvec}_{\bot}(\xvec) /M_0$ 
have very small average value \cite{Holstein_Primakoff_1940}. In this respect, we want to
derive the expansion of the Hamiltonian \eqref{eq:magnetization_hamiltonian}  around the 
classical ground state in powers of the operator $\hat{\Mvec}_{\bot}(\xvec) /M_0$. 
To this aim, we first express, 
by using eq.\eqref{eq:commu_rels_on_M_b} and eq.\eqref{eq:Mvec_i_decomp}, 
the operators	$\hat{M}^{03}(\xvec)$ 
in terms of the operator $\hat{\Mvec}_{\bot}(\xvec)$ as follows
    \begin{align}	
    		\label{eq:M03_expansion_over_GS} 
    \frac{\hat{M}^{03}(\xvec)}{M_0} =  \left[1 - \frac{(\hat{\Mvec}_{\bot}(\xvec))^2 }{2M_0^2} \right] + 	 {F}\left(\frac{\hat{\Mvec}_{\bot}(\xvec)}{M_0}\right)
    \end{align}
    where $F(u)= \sqrt{1-u^2}-1+u^2/2$ 
    is of order $\mathcal{O}(u^4)$ and 
     we have considered 
the approximation    $	M_0^2 \left(1+1/S \right) \approx M_0 ^2$.

   By substituting eqs.\eqref{eq:Mvec_i_decomp},\eqref{eq:M03_expansion_over_GS} into 
   the Hamiltonian \eqref{eq:magnetization_hamiltonian},
   we can express  $\hat{H}$   as function of $\hat{\Mvec}_{\bot}$
   and decompose it in terms that depend on increasing powers of $\hat{\Mvec}_{\bot}(\xvec)$:
   \begin{align}   	\label{eq:expansion_of_H_M}
   	\hat{H} =  	H[{{\Mvec}_0} ]  
   	+	H_1[{\hat{\Mvec}}_{\bot}] +
  	H_2[{\hat{\Mvec}}_{\bot}] +   	H_{\ge 3}[{\hat{\Mvec}}_{\bot}] \, ,
   \end{align}
 where $H_k[{\hat{\Mvec}}_{\bot}]$ contains power of order $k$
 of ${\hat{\Mvec}}_{\bot}(\xvec)$. The first order term vanishes as a consequence
 of the fact that $\Mvec_0$ is a classical micromagnetic equilibrium (see Appendix \Ref{appendix:expansion_of_H_1}).
When considering the limit of large $S$, the
normalized quantities $H_k[{\hat{\Mvec}}_{\bot}]/(\mu_0 M_0^2/2)$,
are infinitesimal of order $k$ in ${\hat{\Mvec}}_{\bot}/M_0$ (see  Appendix \Ref{appendix:expansion_of_H_1})
and this implies that the dominant nonconstant and nonzero term in the expansion \eqref{eq:expansion_of_H_M}
is the quadratic term:
\begin{widetext}
	\begin{align}
		\nonumber
		&  \hat{H}_2 =	H_2[{\hat{\Mvec}_{\bot}} ]=   \int_\Omega  dV  \left\{ 
		\frac{A_{kh}^{uv}}{M_0^2}\, \frac{\partial \hat{M}_\bot^u}{\partial x_k}\frac{\partial \hat{M}_\bot^{v} }{\partial x_h}	  
		+ \frac{1}{2}{\frac{D_{k}^{uv}}{M_0^2}}\left(  \hat{M}_\bot^u \frac{\partial \hat{M}_\bot^v}{\partial x_k} 
		- \frac{\partial \hat{M}_\bot^u}{\partial x_k} \hat{M}_\bot^v \right)  +   
		\hat{M}^u_\bot \frac{K^{uv}}{M_0^2} \hat{M}^v_\bot + \right. \\
		\label{eq:Quadratic_Hamiltonian_M}
		&  \qquad \qquad \qquad \qquad \qquad  \qquad \qquad \qquad \qquad \qquad \qquad \qquad  \qquad
		\left. 	- \frac{\mu_0}{2}\hat{\Mvec }_\bot\cdot \Hvecm[\hat{\Mvec}_\bot] + \frac{ \mu_0 }{2} \lambda_0 (\hat{\Mvec}_\bot)^2 	\right\} \, ,
	\end{align}
\end{widetext}
where $\lambda_0=\lambda_0(\xvec)$ is the scalar field defined in eq.\eqref{eq:Brown_equilibrium}.

The operator \eqref{eq:Quadratic_Hamiltonian_M} describes the quantum  Hamiltonian associated
to low energy quantum magnetization dynamics around the classical ground state, in the limit of large $S$.
When the operator \eqref{eq:Quadratic_Hamiltonian_M} is computed on the classical field $\Mvec_\bot(\xvec)$, pointwise orthogonal to
$\mvec_0(\xvec)$, it gives the Hessian functional\cite{dAquino_2009} of the classical Hamiltonian $H[\Mvec]$ at the ground state $\Mvec_0$. For
this reason $H_2[\Mvec_\bot]$ is a positive definite functional.

	\section{Linear Quantum Magnetization Dynamics}
	\label{sec:Quantum Micromagnetic Spin-waves}
	
	In this section, we derive the linearized equation governing the evolution
	of the quantum field  $\hat{\Mvec}_\bot$ in the Heisenberg picture.
	Since $\hat{\Mvec}_\bot$ is function of the operators
	$\hat{M}^{01}$, $\hat{M}^{02}$   (see eq.\eqref{eq:Mbot_eo1_eo2}), the evolution of  $\hat{\Mvec}_\bot$
	is related to the evolution of these operators that is governed by the equations
	\begin{align}
		\label{eq:Heisenberg_pic_1_01}
     i\hbar \frac{\partial}{\partial t} \hat{M}^{0a}(\xvec,t) = 
		\left[\hat{M}^{0a}(\xvec,t) \, , \,    \hat{H} \right] \, ,
	\end{align}
for $a=1,2$. In order to keep  terms that are linear in $\hat{\Mvec}_\bot$ at the right-hand-side of eq.\eqref{eq:Heisenberg_pic_1_01}, 
we assume the following approximated commutation relations \cite{Kittel_book_1963, Mills_2006}
	\begin{align}
		\label{eq:commu_rels_on_M_a_0_tris}
	&	[\hat{M}^{01}(\xvec), \hat{M}^{02}(\xvec^\prime)] \approx
		-i ({\gamma \hbar})M_0 \hat{\mathbbm{1}} \delta(\xvec-\xvec^\prime) \, ,  \\
			\label{eq:third_approx_comm_rel}
	&  	\hat{M}^{03}(\xvec,t) \approx M_0 \hat{\mathbbm{1}} \, .
	\end{align}
The other approximation to arrive to the linearized equation of motion is to keep in eq.\eqref{eq:Heisenberg_pic_1_01}
only  the quadratic part of the Hamiltonian given by eq.\eqref{eq:Quadratic_Hamiltonian_M}. 
This leads to the following linear equation of motion:  
	\begin{align}
		\label{eq:Heisenberg_M_linearized}
		i\hbar \frac{\partial} {\partial t} \hat{M}_\bot^s(\xvec,t)  = \left[\hat{M}_\bot^s(\xvec,t) \, , \,   
		\hat{H}_2 \right] \, ,
	\end{align}
	with $s=1,2,3$, where the commutators at the right-hand-side of eq.\eqref{eq:Heisenberg_M_linearized}
	are computed by taking into account the approximate commutation 
	relations \eqref{eq:commu_rels_on_M_a_0_tris}-\eqref{eq:third_approx_comm_rel}.
	In this respect, from eqs.\eqref{eq:commu_rels_on_M_a_0_tris}-\eqref{eq:third_approx_comm_rel} follows that 
	\begin{align}
		\label{eq:Mbot_comm_rel_cart}
		[\hat{M}_\bot^s(\xvec),\hat{M}_\bot^u(\xvec^\prime) ]
		\approx i({\gamma \hbar})M_0   \hat{\mathbbm{1}}  	\Lambda_0^{su}(\xvec)  \delta(\xvec-\xvec^\prime) \, ,
	\end{align}
    (see Appendix \ref{appendix:expansion_of_H_2}) where $\Lambda_0^{su}(\xvec)$, with $u,s=1,2,3$, are the component of the second order
    tensor  $\Lambda_0(\xvec)$ with the following property
    \begin{align}
    	\label{eq:Lambda_0_def}
    	\Lambda_0(\xvec) \cdot \vvec(\xvec) = \mvec_0(\xvec) \times \vvec(\xvec) \, ,
    \end{align}
   where $\vvec(\xvec)$ is a generic vector field.

By using the expression \eqref{eq:Quadratic_Hamiltonian_M}
and the commutation relations \eqref{eq:Mbot_comm_rel_cart}
in eq.\eqref{eq:Heisenberg_M_linearized}, after some algebra
(see Appendix \ref{appendix:expansion_of_H}), it is possible to 
derive the explicit form of the equation of motion 
    \begin{align}
    	\label{eq:LL_equation_spin-waves}
    	\frac{\partial} {\partial t} \hat{\Mvec}_\bot(\xvec,t)  = \gamma\mu_0 M_0 \mvec_0(\xvec) \times  \Dvec_{0\bot}[\hat{\Mvec}_\bot] \, ,
    \end{align}
 where
    \begin{align}
    	\label{eq:def_D0_bot_a}
    		\Dvec_{0\bot}[\hat{\Mvec}_\bot] = \mathcal{P}_\bot(\xvec) \cdot \Dvec[\hat{\Mvec}_\bot] + \lambda_0 \hat{\Mvec}_\bot \, ,
    \end{align}
with the operator $\Dvec[\,\cdot\,]$  given in eq.\eqref{eq:Dvec_operator}, and 
where $\mathcal{P}_\bot(\xvec)$ is the second order tensor of components
\begin{align}
	\label{eq:local_projection}
	\mathcal{P}^{uv}_\bot(\xvec) = e^u_{01}(\xvec)e^v_{01}(\xvec)+e^u_{02}(\xvec) e^v_{02}(\xvec) \, ,
\end{align}
with $u,v=1,2,3$, that pointwise projects vector fields in the plane orthogonal to $\mvec_0(\xvec)$.
Equation \eqref{eq:LL_equation_spin-waves} is the quantum version 
of the Landau-Lifshitz equation linearized around the ground state $\Mvec_0$.

In the derivation of eq.\eqref{eq:LL_equation_spin-waves}, we use the following boundary 
conditions for  $\hat{M}^u_\bot(\xvec)$: 
    \begin{align}
    	\label{eq:bc_m_perp}
    	n_k \left({2 A_{kh}^{uv}} \frac{\partial \,\,}{\partial x_h} + n_k { D_{k}^{uv}}\right) \hat{M}_\bot^u(\xvec) = 0  \qquad \text{on} \,\,\, \partial \Omega  \, ,
    \end{align}
 with $v=1,2,3$. 
The important consequence of these boundary conditions is the following
reciprocity property (self-adjointness) of the operator $\Dvec_{0\bot}[\,\cdot\,]$
\begin{align}
	\nonumber
	&\int_\Omega \!\!dV \left\{ \hat{\Fvec}_\bot(\xvec)\cdot  \, \Dvec_{0\bot}[\hat{\Gvec}_\bot]    \right\} = \\
	\label{eq:reciprocity_D0_bot}
	&  \qquad \qquad \int_\Omega\!\! dV \left\{ \Dvec_{0\bot}[\hat{\Fvec}_\bot]\cdot \hat{\Gvec}_\bot(\xvec)]    \right\} \, ,
\end{align}
where $\hat{\Fvec}_\bot(\xvec)$ and $\hat{\Gvec}_\bot(\xvec)$ are fields, possibly 
with noncommuting cartesian  components, orthogonal to $\mvec_0(\xvec)$ and 
satisfying boundary conditions \eqref{eq:bc_m_perp}. The identity \eqref{eq:reciprocity_D0_bot} is
obtained by using boundary conditions \eqref{eq:bc_m_perp}, integration by parts, the reciprocity (self-adjointness) 
of the magnetostatic field operator and symmetry of the tensor $K^{uv}$ (see Sec.\,\ref{appendix:expansion_of_H}). 
This identity is essential in the derivation of the torque-like form of eq.\eqref{eq:LL_equation_spin-waves}
(see Appendix\,\ref{appendix:expansion_of_H}).
 
The boundary conditions \eqref{eq:bc_m_perp} are an extension of those used in classical micromagnetic theory 
to the quantum formulation.  These conditions, and the reciprocity theorem \eqref{eq:reciprocity_D0_bot} that follows from them,
are essential for ensuring the formal consistency of the micromagnetic 
quantum field description presented here with the quantum theory based on the 
quantization of classical harmonic oscillators associated with the amplitude of classical normal modes\,\cite{Mills_2006},\cite{Mills_2007}.

By using the self-adjoint nature of the operator $\Dvec_{0\bot}[\,\cdot\,]$  expressed
in eq.\eqref{eq:reciprocity_D0_bot},  the quadratic Hamiltonian \eqref{eq:magnetization_hamiltonian}  
can be written in  the following form
 \begin{align}
\label{eq:quadratic_Ham_D0_form} 
\hat{H}_2=\int_\Omega dV \left\{   \frac{\mu_0}{2}	\hat{\Mvec}_\bot(\xvec) \cdot \Dvec_{0\bot}[ 	\hat{\Mvec}_\bot]    \right\} 	\,.
 \end{align}
From this last expression, together with the positive definite nature of the quadratic Hamiltonian, 
we conclude that the operator $\Dvec_{0\bot}[\,\cdot\,]$ is positive definite.

In conclusion of this section, we underline that we have 
formulated the linearized Landau-Lifshitz \eqref{eq:LL_equation_spin-waves} and
the associated Hamiltonian operator  \eqref{eq:quadratic_Ham_D0_form}, 
in terms of the field $\hat{\Mvec}_\bot(\xvec)$ and the operator  $\Dvec_{0\bot}[\, \cdot \,]$, imposing the boundary conditions
\eqref{eq:bc_m_perp} on the cartesian components   $\hat{M}^u_\bot(\xvec)$.  
 By taking into account 
eq.\eqref{eq:Mbot_eo1_eo2}, the theory can be alternatively formulated
in terms of the two operators $\hat{M}^{01}(\xvec), \hat{M}^{02}(\xvec)$.
Although this latter formulation would be mathematically equivalent to the one presented,
the boundary conditions for the fields  $\hat{M}^{01}(\xvec), \hat{M}^{02}(\xvec)$,
that are given by 
\begin{align}
	\nonumber
	&	\left( n_k {2 A_{kh}^{uv}} \frac{\partial }{\partial x_h} + n_k { D_{k}^{uv}} \right) \cdot\\
	\label{eq:bc_M0a}
	&   \quad\cdot\left(	e_{01}^u(\xvec)  \hat{M}^{01}(\xvec) +  e_{02}^u(\xvec)  \hat{M}^{02}(\xvec)  \right) = 0 
	\,\,\, \text{on} \,\, \partial \Omega  \, ,
\end{align}
are considerably more convoluted with respect to the  the boundary conditions \eqref{eq:bc_m_perp}. 
The central role that boundary conditions play in the micromagnetics of nanostructures suggests 
that formulation based on the Cartesian components of $\hat{M}^u_\bot(\xvec)$ is, in this respect, 
preferable.
    
\section{Diagonalization of Quantum Magnetization Dynamics}
\label{sec:Diagonalization of Quantum Magnetization Dynamics}

In order to introduce magnons, as quasi-particles corresponding
to excitations of magnetization quantum field,  we have to express 
$\hat{\Mvec}_\bot$ in terms of the appropriate creation-annihilation operator.
To this aim, we consider a canonical transformation that connects the quantum field 
$\hat{\Mvec}_\bot$ with the set of annihilation-creation operators
$(\hat{c}_m,\hat{c}^\dag_m)$, with $m \in \mathbb{N}^*$ (nonzero positive integers),
that has the form, in the Heisenberg picture, of the following generalized Fourier expansion:
\begin{align}
	\nonumber
&	\hat{\Mvec}_\bot(\xvec,t) =  \\
\label{eq:canonical_trasf_diag_H}
&  \qquad   M_0\sum_{m \in \mathbb{N}^* } \left\{ \hat{c}_m(t) \varphivec_m(\xvec) +  \hat{c}_m^\dag(t) \varphivec_m^*(\xvec) \right\} \, ,
\end{align}
where  $\varphivec_m(\xvec)\cdot\mvec_0(\xvec)=0$, and where
$(\hat{c}_m(t)$, $\hat{c}_m^\dag(t))$ are the annihilation-creation operators
in the Heisenberg picture. The operators $(\hat{c}_m,\hat{c}^\dag_m)$ in the
Schroedinger picture corrispond to the values of $\hat{c}_m(t)$, $\hat{c}_m^\dag(t)$
for $t=0$.

We notice that, with the above normalizations, both the operators $\hat{c}_m$, $\hat{c}_m^\dag$ and
the functions $\varphivec_m(\xvec)$ are dimensionless quantities. 
In order to simplify algebraic manipulations, we can write
\eqref{eq:canonical_trasf_diag_H} as a single summation as follows
\begin{align}
	\label{eq:canonical_trasf_diag_H_Z}
\hat{\Mvec}_\bot(\xvec,t)	= M_0 \sum_{p \in \mathbb{Z}^*} \hat{c}_p(t) \varphivec_p(\xvec)  \, ,
\end{align}
where $\mathbb{Z}^*$ is the set of nonzero integers, and we assume that
\begin{align}
\label{eq:def_p_q_eigenquant}
	\hat{c}_{-p}  = \hat{c}_{p}^\dag \, , \quad \varphivec_{-p}(\xvec)= \varphivec^*_p(\xvec) \, .
\end{align}
The transformation \eqref{eq:canonical_trasf_diag_H_Z} must be such to satisfy the  two conditions
discussed in the sequel.

The first condition is that, under \eqref{eq:canonical_trasf_diag_H_Z}, the Hamiltonian operator
\eqref{eq:quadratic_Ham_D0_form} is reduced to the  diagonal form 
\begin{align}
	\label{eq:cond_diag_H_canonical_c}
	\hat{H}_2
	=   \sum_{m \in \mathbb{N}^* }\frac{ \hbar \omega_m}{2} \left( \hat{c}_m^\dag \hat{c}_m + \hat{c}_m \hat{c}_m^\dag \right)
	=   \sum_{p \in \mathbb{Z}^*}\frac{ \hbar \omega_p}{2}{\hat{c}_p}^\dag \hat{c}^{}_p \, ,
\end{align}
corresponding to a denumerable set of noninteracting quantum harmonic oscillators.
In eq.\eqref{eq:cond_diag_H_canonical_c}, $\omega_{p} = \omega_{-p}$, for $p\in \mathbb{Z}^*$, are real positive quantities that have the physical 
meaning of angular frequencies and have  to be determined together with the $\varphivec_p(\xvec)$.  
By substituting eq.\eqref{eq:canonical_trasf_diag_H_Z} into the expression \eqref{eq:Quadratic_Hamiltonian_M}, and imposing
the equality with the diagonalized Hamiltonian \eqref{eq:cond_diag_H_canonical_c},
we obtain that the basis vector fields $\varphivec_p(\xvec)$ have to satisfy the following 
orthogonality conditions
\begin{align}
\label{eq:D0_orthogonality_b}
	\int_\Omega dV \left\{   	\varphivec_p(\xvec)^* \cdot \Dvec_{0\bot }[ \varphivec_q]    \right\}= \frac{\hbar \omega_p}{\mu_0 M_0^2 } \delta_{pq} \, ,
\end{align}
where $p,q \in \mathbb{Z}^*$, and the second definition in eq.\eqref{eq:def_p_q_eigenquant} is taken into account. Given the property of $\Dvec_{0\bot}[\,\cdot\,]$ of being  real positive definite self-adjoint operator,   
the integrals in eq.\eqref{eq:D0_orthogonality_b} must be positive real quantities confirming the fact that  all the $\omega_p$ are positive.

The second condition on the  transformation \eqref{eq:canonical_trasf_diag_H_Z} is that 
it must be such to produce the following canonical commutation relations:
\begin{align}
	\label{eq:creation_annihilation_comm_rels}
	[\hat{c}_p,\hat{c}^\dag_q] =\tilde{\delta}_{pq}  \hat{\mathbbm{1}} \, , \quad 
\end{align}
for $p,q \in \mathbb{Z}^*$, where
$$\tilde{\delta}_{pq}= \delta_{pq} \text{sgn}(p)$$ 
and $\text{sgn}(p)$ is the signum of $p$, provided that $\hat{\Mvec}_\bot(\xvec)$ satisfies the commutation relations \eqref{eq:Mbot_comm_rel_cart}.
By substituting eq.\eqref{eq:canonical_trasf_diag_H_Z} into 
eq.\eqref{eq:Mbot_comm_rel_cart}, we obtain
\begin{align}
	\nonumber
&	\sum_{p,q \in \mathbb{Z}^*} 	[\hat{c}_q,\hat{c}^\dag_p]  \varphi_q^s(\xvec) \left( \varphi_p^u(\xvec^\prime) \right)^*=	\\
&	\qquad\quad =  i\frac{({\gamma \hbar})}{M_0}   \hat{\mathbbm{1}}  	\Lambda_0^{su}(\xvec^\prime)  \delta(\xvec-\xvec^\prime) \, .
\end{align}
Then, we multiply both sides of the last equation by the quantity
\begin{align}
	\varphi_l^r(\xvec^\prime) 	\Lambda_0^{ur}(\xvec^\prime)
\end{align}
with $l\in \mathbb{Z}^*$, and integrate with respect to $\xvec^\prime$, obtaining
\begin{align}
	\nonumber
&	\sum_{p,q \in \mathbb{Z}^*} 	[\hat{c}_q,\hat{c}^\dag_p]  \varphi_q^s(\xvec)
	\int_\Omega dV^\prime \left\{	\varphi_l^r(\xvec^\prime) 	\Lambda_0^{ur}(\xvec^\prime) \left( \varphi_p^u(\xvec^\prime) \right)^*	\right\}\\
	\label{eq:normalization_conditions_seed}
&	\quad =-  i\frac{({\gamma \hbar})}{M_0}   \hat{\mathbbm{1}}  	\delta^{sr} 	\varphi_l^r(\xvec) 
	=-  i\frac{({\gamma \hbar})}{M_0}   \hat{\mathbbm{1}}  	\varphi_l^s(\xvec)  \, .
\end{align}
Now, if we impose the  condition
\begin{align}
	\label{eq:normalization_con_eigenfunctions}
	\int_\Omega dV^\prime \left\{	\varphi_l^r(\xvec^\prime) 	\Lambda_0^{ur}(\xvec^\prime) \left( \varphi_p^u(\xvec^\prime) \right)^*	\right\}
	= -  i\frac{({\gamma \hbar})}{M_0}\tilde{\delta}_{lp} \, , 
\end{align}
in eq.\eqref{eq:normalization_conditions_seed}, we obtain the relations  \eqref{eq:creation_annihilation_comm_rels}. 
The normalization condition \eqref{eq:normalization_con_eigenfunctions} can be written   in vector form as
\begin{align}
	\label{eq:normalization_vector_form}
	\int_\Omega dV^\prime \left\{	\varphivec_p^*(\xvec^\prime) \cdot \mvec_0(\xvec)\times \varphivec_l(\xvec^\prime) 	\right\}
	= -  i\frac{({\gamma \hbar})}{M_0}\tilde{\delta}_{lp} \, . 
\end{align}
This last equation is the generalization of the normalization conditions reported in the paper of Walker\cite{Walker_1957} for the
case of a sample subject to magnetostatic interactions and with a spatially uniform ground state (see eq.(B3) in Ref.\,\cite{Walker_1957}). 
These normalization conditions were extended, for spatially uniform ground states, to the case where isotropic exchange interactions are also present in the 
paper of  Mills\cite{Mills_2006}. The derivation above extend these normalization conditions to the much more general case of noncollinear  ground state. 
In addition, in our formalism, we also include anisotropic and DM exchange interactions.

\begin{figure*}[!t]
	\centering 
	\includegraphics[width=16cm]{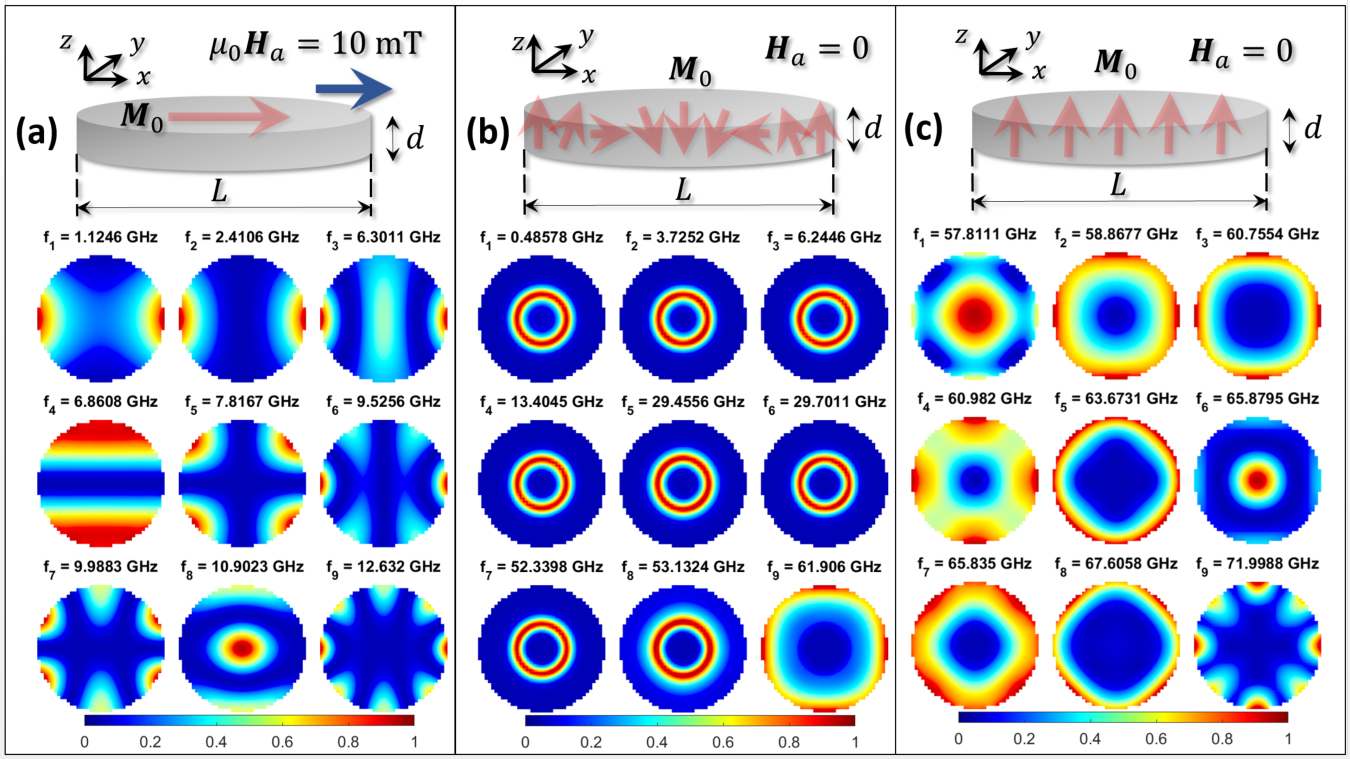} 
	\caption{Computed normal mode spatial profiles for thin-disks. Each panel reports the first nine eigenmodes $\bm \varphi_m(\bm x), m=1,\ldots,9$ (in ascending order of eigenfrequency $f_m$) computed around the equilibrium $\bm M_0$. The color map refers to the amplitude $|\bm \varphi_m(\bm x)|$ (normalized to the maximum value, blue for minimum and red for maximum). (a) thin-disk with $L=100$ nm, $d=5$ nm, magnetized along $x$ and subject to an external field $\mu_0 H_a=10$ mT; (b) thin-disk with $L=100$ nm, $d=0.4$ nm, with a Bloch Skyrmion ground state $\bm M_0$, no applied field; (c) thin-disk with $L=100$ nm, $d=0.4$ nm, magnetized along $z$, no applied field. Details on the material parameters used in the calculations are reported in Sec.\ref{sec:applications}.}
	\label{fig:computed normal modes disk 100nm}
\end{figure*}

To complete the discussion we have to illustrate the procedure to determine the set of functions
$\varphivec_p(\xvec)$ and the corresponding set of positive scalar quantities $\omega_p$ for 
$p\in \mathbbm{Z}^*$. In this respect, by combining eqs. \eqref{eq:LL_equation_spin-waves}, \eqref{eq:canonical_trasf_diag_H} and \eqref{eq:normalization_vector_form}, we notice that the desired quantities can be obtained 
by solving the following generalized eigenvalue problem
\begin{align}
	\label{eq:Gen_EigenProblem_LL}
  \Dvec_{0\bot}[ \varphivec_p](\xvec) = \nu_p  i \Lambda_0(\xvec) \cdot  \varphivec_p(\xvec) \quad \text{in} \,\,  \Omega
\end{align}
with boundary conditions
	\begin{align}
	\label{eq:bc_phi_perp}
	n_k \left( {2 A_{kh}^{uv}} \frac{\partial }{\partial x_h} +  { D_{k}^{uv}} \right) \varphi_p^u(\xvec) = 0  \quad \text{on} \,\, \partial \Omega  \, ,
\end{align}
that arises in the study of micromagnetic classical normal modes\cite{dAquino_2009}
in finite structures. Indeed, the problem \eqref{eq:Gen_EigenProblem_LL}, can be readily obtained
searching for time-harmonic solutions of the type
$\varphivec(\xvec)e^{-i\omega t}$ for the classical counterpart of the eq.\eqref{eq:LL_equation_spin-waves}. By  using the fact that, on vector fields locally orthogonal to $\mvec_0(\xvec) $
and satisfying boundary conditions \eqref{eq:bc_phi_perp}, 
$\Dvec_{0\bot}[ \,\cdot\,] $ is real positive definite and self-adjoint operator while  
$i \Lambda_0(\xvec) $ is an Hermitian operator, one can prove that eigenvalues are
real and such that\cite{dAquino_2009} 
\begin{align}
\nu_m >0 \, , \,\, m \in \mathbb{N}^* \, , \qquad 	\nu_{-p} = -\nu_p \, , \,\, p \in \mathbb{Z}^* \, .
\end{align}
In addition, eigenfunctions $\varphivec_k(\xvec)$ can be normalized in such a way to satisfy both
conditions \eqref{eq:D0_orthogonality_b},\eqref{eq:normalization_vector_form}. 
The sequence $\nu_p$ determines the
angular frequencies $\omega_p$ of harmonic oscillators contained in the diagonalized Hamiltonian 
through the equation
\begin{align}
	\label{eq:omega_nu_rel}
	\omega_m = (\gamma \mu_0 M_0) \nu_m\, , \quad m \in \mathbb{N}^* \, ,
\end{align}
These are the angular frequencies of time-harmonic evolution:
\begin{align}
\label{eq:evo_anni-creation_ops}
    \hat{c}_m(t) = \hat{c}_m(0) e^{-i\omega_m t} \, , \,\, 
    \hat{c}^\dag_m(t) = \hat{c}^\dag_m(0) e^{i\omega_m t} \, ,
\end{align}
of annihilation-creation operators.
Substitution of eq.\eqref{eq:evo_anni-creation_ops} into eq.\eqref{eq:canonical_trasf_diag_H},
leads to the time evolution
of $\hat{\Mvec}(\xvec,t)$.
The general proofs of the properties of eigenvalues and eigenfunctions stated 
previously are contained in Ref.\cite{dAquino_2009}.
The extension of these proofs to the case treated here that includes DM interaction (i.e. $D^{uv}_k\neq 0$) and anisotropic exchange  is possible owing to 
the self-adjoint and the positive definite nature of 
the operator $ \Dvec_{0\bot}[\,\cdot\,]$.
These formal properties
provide a general framework   for approaching the diagonalization problem
in the most general case.

After presenting the general method based on normal mode analysis, it is important 
to briefly connect this approach to the traditional plane-wave method. For a 
body with large dimensions, the system can be approximated as having infinite 
spatial extent. If, furthermore, the magnetic material is spatially homogeneous, 
the normal mode analysis reduces to the traditional plane-wave approach \cite{Sparks_book} 
in which the expansion \eqref{eq:canonical_trasf_diag_H}
becomes
\begin{align}
	\nonumber
	&\hat{\Mvec}_\bot(\xvec,t) = \\
    \label{eq:plane_waves_diag_H_1}
    & \quad \,  M_0\sum_{\kvec } \left\{ \hat{c}_\kvec(t) \varphivec(\kvec)e^{i\kvec\cdot \xvec} +   \hat{c}_\kvec^\dag(t) \varphivec^*(\kvec)e^{-i\kvec\cdot \xvec}  \right\} \, ,
\end{align}
where the wave vector $\kvec$ has here also to role of indexing modes, and the vectors $\varphivec(\kvec)$ have
to satisfy a Walker's type normalization conditon analogous to \eqref{eq:normalization_vector_form}.
The annihilation-creation operators evolve according to the equation
\begin{align}
    \hat{c}_\kvec(t) = \hat{c}_\kvec(0)e^{-i\omega(\kvec)t} \, , \,\,  
     \hat{c}^\dag_\kvec(t) = \hat{c}^\dag_\kvec(0)e^{i\omega(\kvec)t} \,,
\end{align}
where the dependence of the angular frequency with respect to $\kvec$ 
is given by the appropriate dispersion relation $\omega = \omega(\kvec)$.
A comparison of the expansions \eqref{eq:canonical_trasf_diag_H},\eqref{eq:plane_waves_diag_H_1}
shows that the role of the normal mode profile vector $\varphivec_m(\xvec)$ is played 
in plane-wave formalims by the vector field $\varphivec(\kvec)e^{i\kvec\cdot \xvec}$. 
In this vector field, the spatial dependence is fully captured 
by the phase factor, while the vector $\varphivec(\kvec)$ describes the position-independent
polarization of  the $\kvec$-th  spin wave. In contrast, for normal mode vectors,
the polarization and the spatial dependence of the mode, also determined by the possibly noncollinear ground state $\bm M_0$, are both encoded in the vector $\varphivec_m(\xvec)$.

To elucidate such circumstance, by using the large-scale methods described in Ref.\cite{dAquino_2009} and implemented in the MaGICo micromagnetic code\cite{MaGICo}, we have computed a set of normal modes $\bm \varphi_m(\bm x)$ for thin disks with similar dimensions starting from different ground states, as reported in Fig.\ref{fig:computed normal modes disk 100nm}. Specifically, panel (a) refers to a disk with diameter $L=100$ nm and thickness $d=5$ nm magnetized in-plane under the action of the external field, panels (b) and (c) refer to a disk with $L=100$ nm and $d=0.4$ nm having a Bloch Skyrmion and quasi-uniform out-of-plane magnetization ground states, respectively.
For instance, one can clearly see in Fig.\ref{fig:computed normal modes disk 100nm}(b) that the spatial amplitude profiles of the normal modes resemble the spatial pattern of the associated Skyrmion ground state that is rotationally-symmetric. Moreover, in the same figure panel we observe that, for the lowest modes, the strongest oscillation amplitude is apparently localized in the region with the steepest spatial gradient of $\bm M_0$.
Further details on the computations are reported in Sec.\ref{sec:applications} where the normal modes profiles are used for the evaluation of thermal equilibrium averages.

\section{Computations of thermal equilibrium averages in thin disks}
\label{sec:applications}

In this section, we want to present sample calculations that demonstrate
the capability of the proposed formalism. Our focus will be on the low-temperature 
thermal equilibrium averages of observables associated with the Cartesian components  
of $\hat{\Mvec}(\xvec)$. Such thermal equilibrium computations are particularly relevant for initializing or engineering magnon populations in magnetic nanostructures,
and this is important for low-temperature applications of quantum magnonics 
in quantum computing and information processing \cite{Yuan2022}, \cite{Yuan2022a}.

In the calculations, we consider the diagonalized
magnon Hamiltonian
\begin{align}
	\label{eq:cond_diag_H_canonical_c_bis}
	\hat{H}_2^\prime 	=   \sum_{m \in \mathbb{N}^* }{ \hbar \omega_m} \hat{c}_m^\dag \hat{c}_m = 
	\sum_{m \in \mathbb{N}^* }{ \hbar \omega_m} \hat{n}_m
\end{align}
where
\begin{equation*}
	\hat{n}_m = \hat{c}_m^\dag \hat{c}_m \, ,
\end{equation*}
is the number operator of magnons with frequency $\omega_m$.
The Hamiltonian \eqref{eq:cond_diag_H_canonical_c_bis} is obtained from \eqref{eq:cond_diag_H_canonical_c} after removing
the zero-point energy. The zero-point contribution in field theory of magnons is actually infinite 
as the spectrum is unbounded from above, similarly to what happens in quantum electrodynamics \cite{Heitler_1936}.
In addition, we are here only concerned with the temperature-dependent contribution to the quantities under investigation, 
which does not include the zero-point energy.

For a generic observable quantity $\hat{F}$, we define the equilibrium  thermal average as
\begin{align}
	\label{eq:thermal_averages}
	\langle \hat{F}  \rangle_T = \text{tr}(\hat{\rho}_\text{eq} \hat{F}) \, ,
\end{align} 
where $\hat{\rho}_\text{eq}$ is the equilibrium density operator given
by 
\begin{align}
	\hat{\rho}_\text{eq} = \frac{1}{Z}\exp(-\beta \hat{H}_2^\prime)
\end{align}
where $\beta = 1/(k_B T)$, $k_B = 1.380649 \cdot 10^{-23}$ J/K is the Boltzmann constant, $T$ the absolute temperature and $Z = \mathrm{tr}(\exp(-\beta \hat{H}_2'))$ is the partition function. 

\begin{figure}[!t]
	\centering
	\includegraphics[width=5cm]{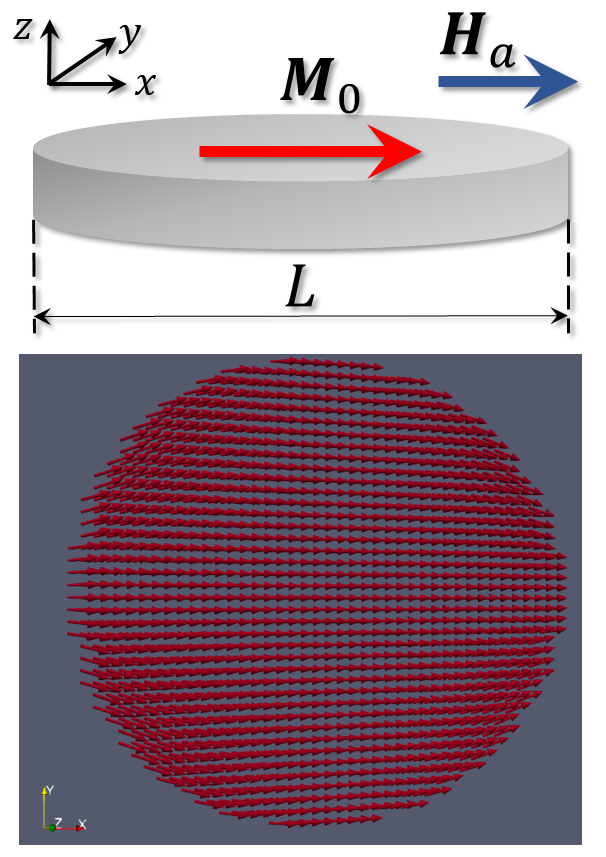}
	\caption{(Top) Thin-disk geometry ($L$ is diameter and $d$ the thickness of the disk), 
		and (bottom) classical micromagnetic in-plane ground state for a soft disk
	subject to an in-plane applied field.}
	\label{fig:Disk_geometry_and_groud_state_uniform}
\end{figure}

\begin{figure}[!t]
	\centering (a) \\
	\includegraphics[width=8cm]{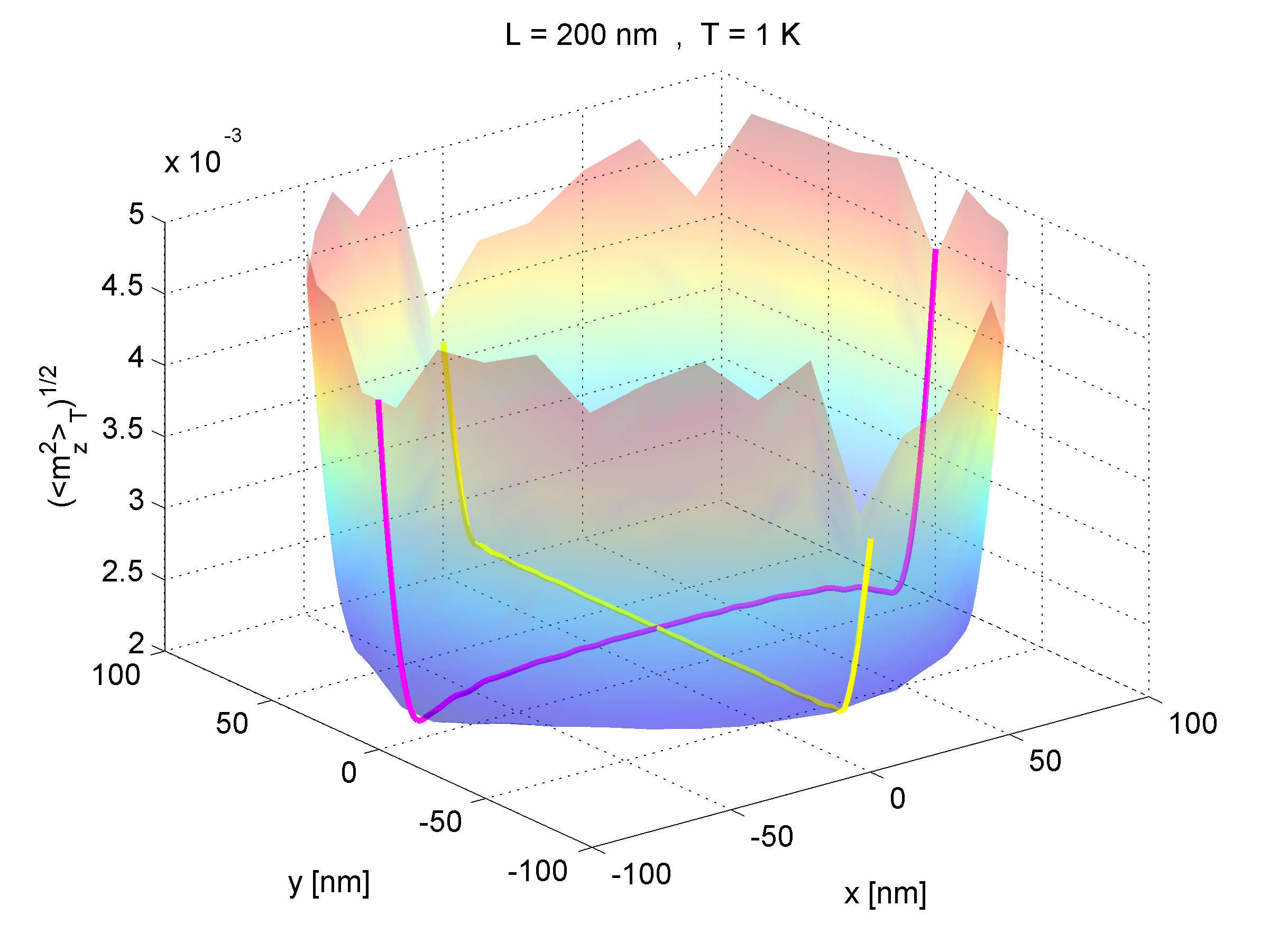} \\ (b) \\
	\includegraphics[width=7cm]{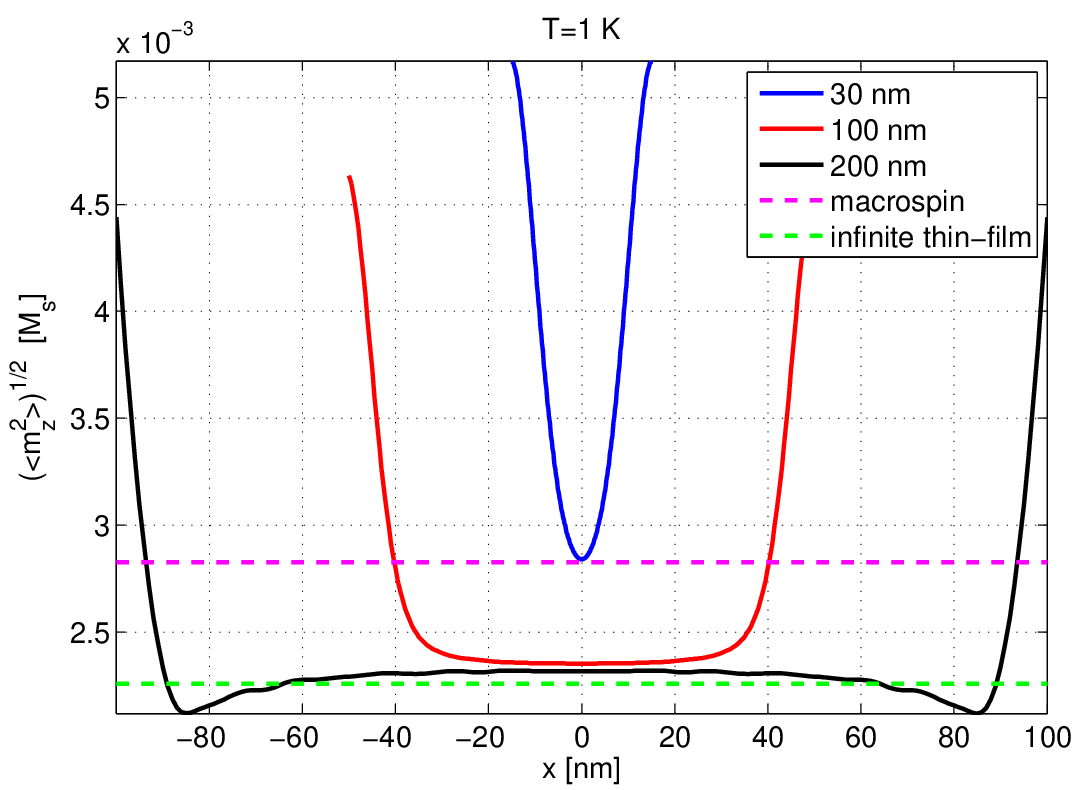}
	\caption{(a) Profile of thermal mean square value of $\hat{M}^z(\xvec)/M_0$ for a thin disk of diameter $L=200$ nm and thickness 5 nm, magnetized along $x$ ($\mu_0 H_a=10$ mT), at $T=1$ K. Magenta (yellow) line refers to a line-cut along the $x$-axis ($y$-axis). (b) line-cut diagram along $x$ of the color map for different dimension $L=30,100,200$ nm compared with the calculation in the infinite thin-film (green) and  macrospin limits (magenta).}
	\label{fig:Disk_profile_mz_vs_T_vs_L}
\end{figure}

By using eq.\eqref{eq:thermal_averages} for the case when 
$\hat{F} = \hat{n}_m$, we obtain 
\begin{align}
		\langle \hat{n}_m \rangle_T  = \frac{1}{e^{\beta\hbar\omega_m}-1} \, , \label{eq:bose statistics average occupation number}
\end{align}
that is the Bose statistics average occupation number.
From the expansion \eqref{eq:canonical_trasf_diag_H}, we can immediately derive
\begin{align}
	\langle	|\hat{M}_\bot^u(\xvec)|^2 \rangle_T  =\sum_{m \in \mathbb{N}^*} 2 \langle \hat{n}_m \rangle |\varphi^u_m(\xvec)|^2 \, . \label{eq:quantum thermal mean square M_perp}
\end{align}
In addition, by using  eq.\eqref{eq:M03_expansion_over_GS}, we can write that
\begin{align}
	\hat{\Mvec}(\xvec) = M_0 \mvec_0(\xvec) + \hat{\Mvec}_\bot - \frac{\hat{\Mvec}_\bot^2}{2M_0}\mvec_0(\xvec)+ \mathcal{O}({\hat{\Mvec}_\bot^3} )
\end{align}
and from this we can derive that 
\begin{align}
\nonumber
&	\langle	|\hat{m}^u(\xvec)|^2 \rangle_T  = |m_0^u(\xvec)|^2 + \\
\label{eq:mean_square_formulas_M_u}
&\sum_{m \in \mathbb{N}^*} 2 \langle \hat{n}_m \rangle_T \left\{|\varphi^u_m(\xvec)|^2-  |\varphivec_m(\xvec)|^2 (m_0^u(\xvec))^2    \right\} \, ,
\end{align}
where we used the notation
\begin{align}
	\hat{m}^u(\xvec) = \hat{M}^u(\xvec)/M_0 \, .
\end{align}
The formulas above are used for the numerical computations outlined in the following.

\begin{figure}[!t]
	\centering (a) \\
	\includegraphics[width=8cm]{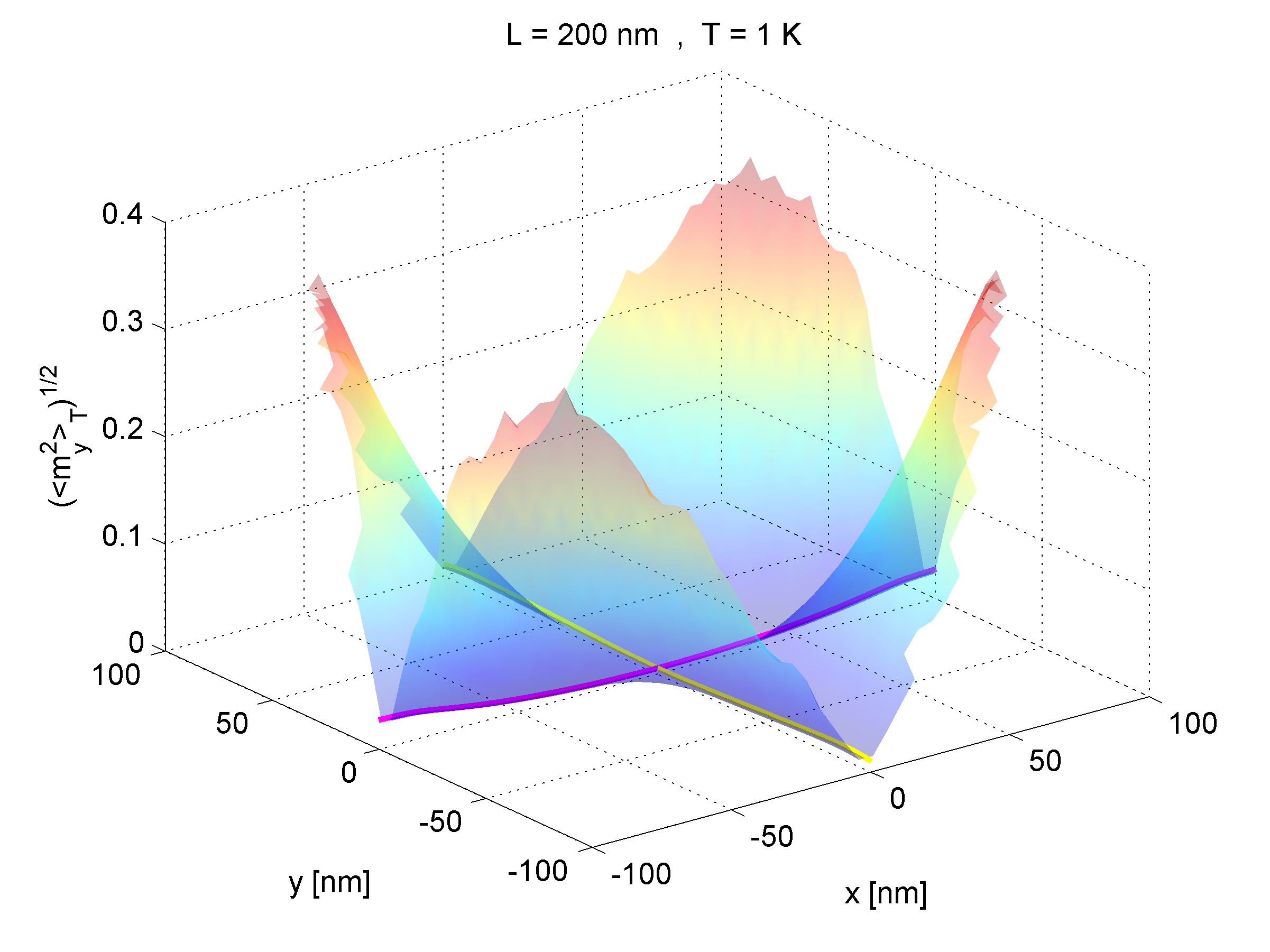} \\ (b) \\
	\includegraphics[width=7cm]{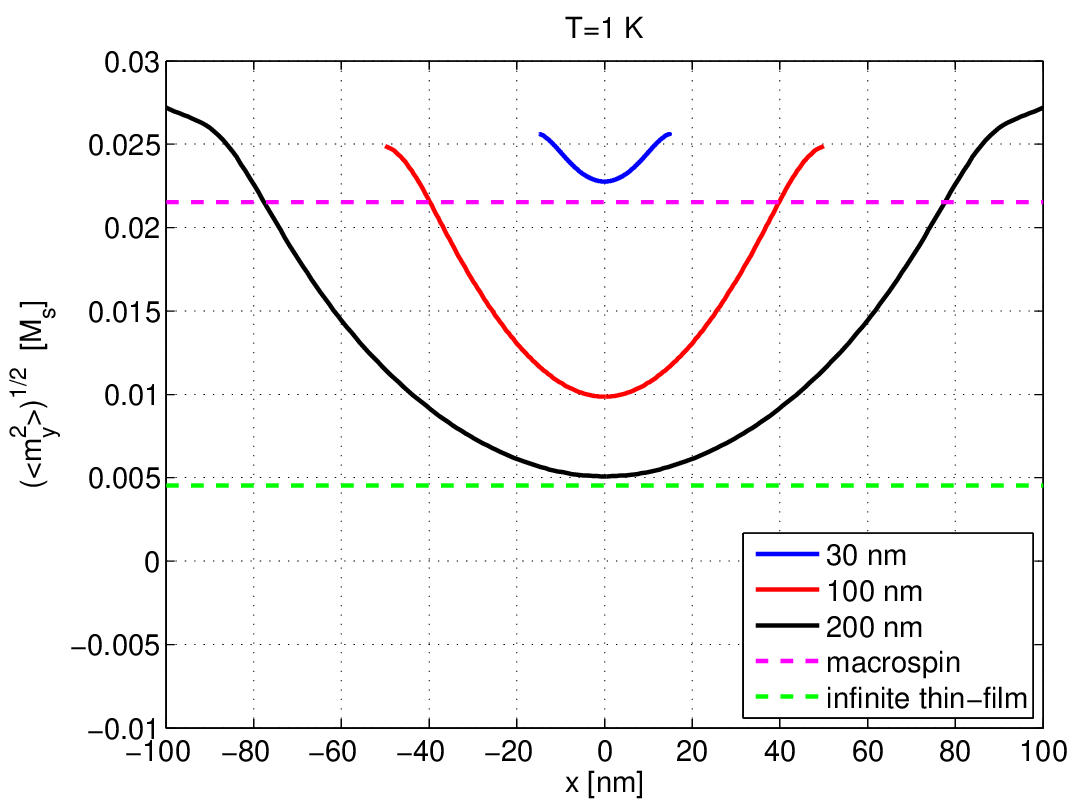}
	\caption{(a) Profile of thermal mean square value of $\hat{M}^y(\xvec)/M_0$ for a thin disk of diameter $L=200$ nm and thickness 5 nm,  magnetized along $x$ ($\mu_0  H_a=10$ mT), at $T=1$ K. Magenta (yellow) line refers to a line-cut along the $x$-axis ($y$-axis). (b) line-cut diagram along $x$ of the color map for different dimension $L=30,100,200$ nm compared with the calculation in the infinite thin-film (green) and  macrospin (magenta) limits.}
	\label{fig:Disk_profile_my_vs_T_vs_L}
\end{figure}

In the remaining part of this section we present numerical computation of thermal 
equilibrium  quantities for thin-disks of submicron spatial scale. We will
consider two cases: a soft thin-disk subject to in-plane saturating applied field,
and a thin-disk made of material with DM interaction and perpendicular anisotropy
that exhibits a Skyrmion ground state. In both cases, the materials have spatially uniform
material parameters, the non-chiral part of the
exchange is isotropic ($A^{uv}_{kh} = A \delta_{uv}\delta_{kh}$) and the
anisotropy tensor  when present is uniaxial ($K^{uv}= K_1 \evec_\text{av}^u\evec_\text{an}^v$ ).

For  both set of numerical computations, the results are compared with those obtained 
in two limiting cases: macrospin limit and infinite thin-film limit. 

In the macrospin limit, the magnetic body is treated under the assumption that the magnetization is spatially uniform. 
The effective field operator \eqref{eq:Dvec_operator} acting on a vector field $\bm F$ simplifies to 
$-D[\bm F]=-D_x F^x\bm e_x-D_y F^y\bm e_y -D_z F^z\bm e_z$, where $(D_x,D_y,D_z)$ are effective demagnetizing factors that account for the 
shape and crystalline anisotropy of the macrospin particle.

In the infinite thin-film limit, the thermal averages are derived analytically using 
the plane-wave formalism (see Appendix\,\ref{Appendix:Quantum spin wave treatment}).
These analytical calculations assume that the magnetization is spatially uniform across the thickness 
of the disk and that the wavevector  $\kvec$  lies in the plane of the disk.

\subsubsection{Thermal equilibrium computations for a soft thin-disk with in-plane ground state.}

We start by considering a thin disk of diameter $L$ and thickness $d=5$ nm as depicted in Fig \ref{fig:Disk_geometry_and_groud_state_uniform}. We assume
that the disk is made of a soft magnetic material with $M_0=800$ kA/m so that anisotropy can be neglected. We take 
the exchange constant with a value $A=13$ pJ/m and no DM interaction. 
We introduce Cartesian axes such that the $z$-axis is normal to the disk plane and the applied field is along the 
$x$-axis. The thickness $d$ of the disk is smaller than the exchange length  $l_{ex} = \sqrt{2A/(\mu_0 M_0^2)}=5.71$ nm and, for this reason, we may safely assume that magnetization field does not depend on the coordinate $z$. The disk is magnetized in-plane along the $x$ axis under the action of a small applied field $\mu_0 H_a=10$ mT.

For the micromagnetic calculation, the structure is discretized in a collection of square prism cells ($1\times 1\times 5$ nm$^3$ when $L=30$ nm, $2.5\times 2.5\times 5$ nm$^3$ for $L=100$ nm and $5\times 5\times 5$ nm$^3$ for $L=200$ nm). The micromagnetic ground state is obtained by relaxing the magnetization starting from an initial condition where the magnetization is aligned with the $x$ axis until the residual (normalized) maximum torque goes below $10^{-5}$. 
Then, the first 200 eigenmodes of the structure (see Fig.\ref{fig:computed normal modes disk 100nm}(a) for the disk with $L=100$ nm) are computed using the large-scale methods described in Ref.\cite{dAquino_2009} and implemented within the MaGICo micromagnetic code\cite{MaGICo}.

\begin{figure}[!t]
	\centering (a) \\
	\includegraphics[width=8cm]{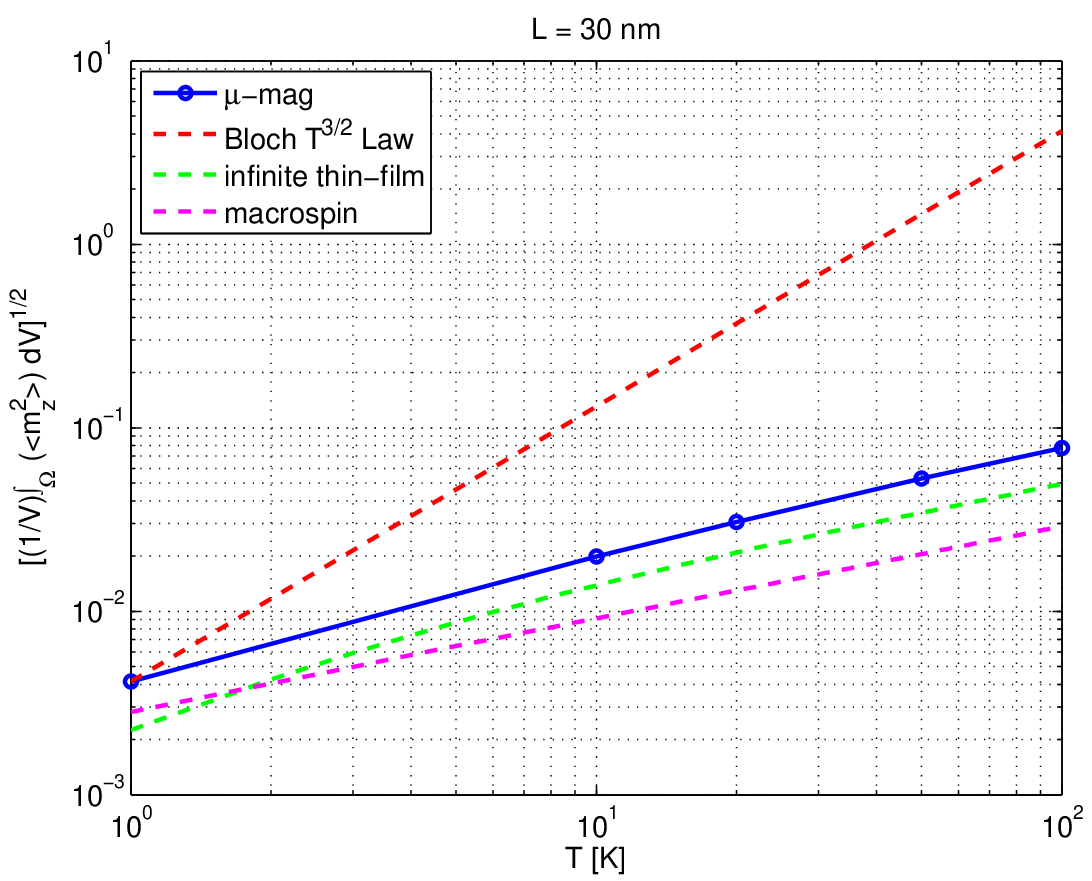} \\ (b) \\
	\includegraphics[width=8cm]{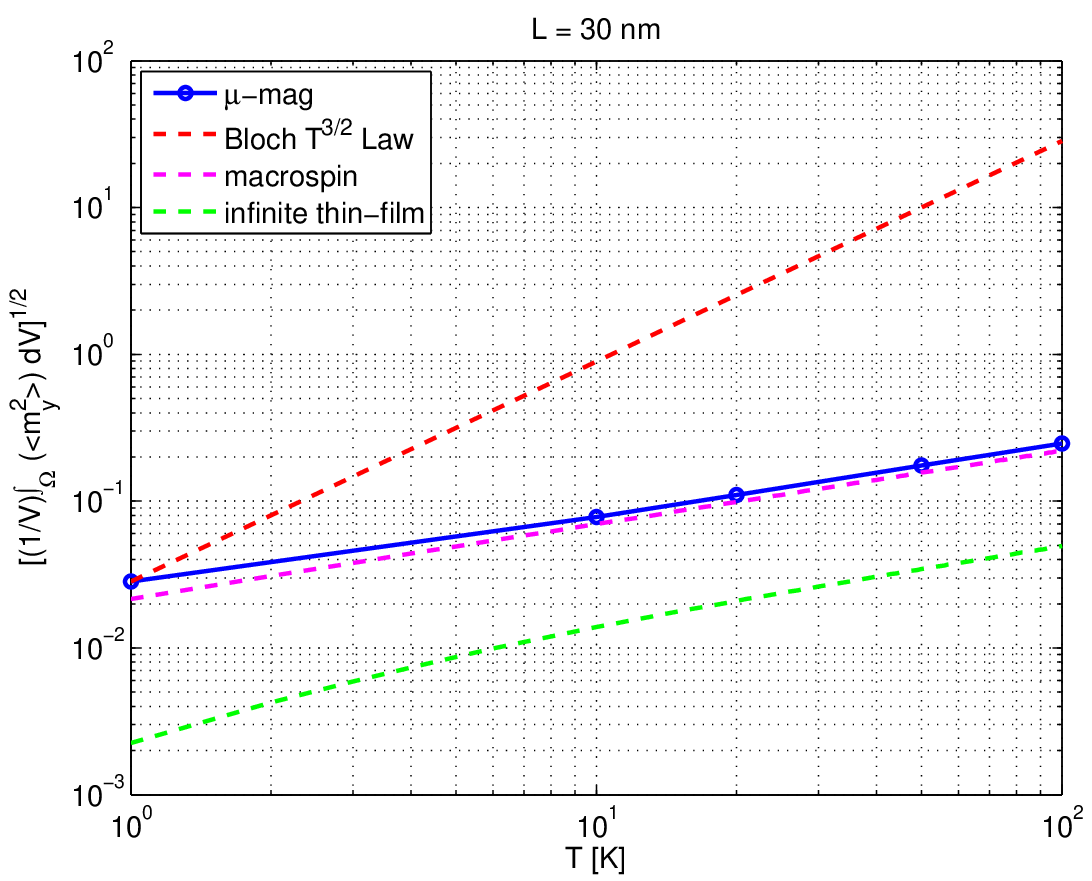}
	\caption{Thermal mean square value of (a) $\hat{M}^z(\xvec)$; (b) $\hat{M}^y(\xvec)$ for the thin disk with thickness 5 nm and diameter $L=30$ nm as function of temperature $T$. Solid blue line refers to full micromagnetic calculation of Eq.\eqref{eq:quantum thermal mean square M_perp}, red dashed line indicates the Bloch's $T^{3/2}$ Law scaling for temperature, green and magenta lines refer to analytical computations in the infinite thin-film and macrospin limits.}
	\label{fig:Disk_thermal_mean_vs_T}
\end{figure}

The results of the calculations for the thermal mean square values of $\hat{M}^z(\xvec)/M_0$ and $\hat{M}^y(\xvec)/M_0$, computed
according to the formula \eqref{eq:mean_square_formulas_M_u}, are reported in Figures \ref{fig:Disk_profile_mz_vs_T_vs_L} 
and \ref{fig:Disk_profile_my_vs_T_vs_L}, respectively. One can see in the upper panels (a) the color maps associated with 
the mean square deviations from the equilibrium along $z$ and $y$ axes for the disk with diameter $L=200$ nm, while a 
line-cut diagram along the $x$ axis at $y=L/2$ is reported in the lower panels (b) for several dimensions $L=30,100,200$ nm. 
The lower panels also report the comparison between micromagnetic calculations (solid blue lines) with analytical calculations 
performed in the infinite thin-film  (see Sec.\ref{Appendix:Quantum spin wave treatment})  and macrospin limits (magenta), showing the consistence of the micromagnetic computations. The effect of the disk boundaries is evident in that the deviation is much more pronounced than in the center. 

In order to investigate the behavior of the thermal mean square value as function of the temperature, the RMS value of $\hat{M}^z(\xvec)/M_0$ and $\hat{M}^y(\xvec)/M_0$ over the disk volume has been computed for several values of $T$. The result is reported in Fig.\ref{fig:Disk_thermal_mean_vs_T}. For the sake of comparison, the Bloch's $T^{3/2}$ Law scaling is reported (red dashed lines) along with the analytical calculation performed in the infinite thin-film and macrospin limits. One can clearly see that the micromagnetic computations are in qualitative agreement with the analytical calculations, but one needs to take into account the spatial inhomogeneity in order to have a quantitatively accurate description of the fluctuation.

\subsubsection{Thermal equilibrium computations for a thin-disk with DM interaction, anisotropy and Skyrmion ground state.}

\begin{figure}[!t]
	\centering
	\includegraphics[width=5cm]{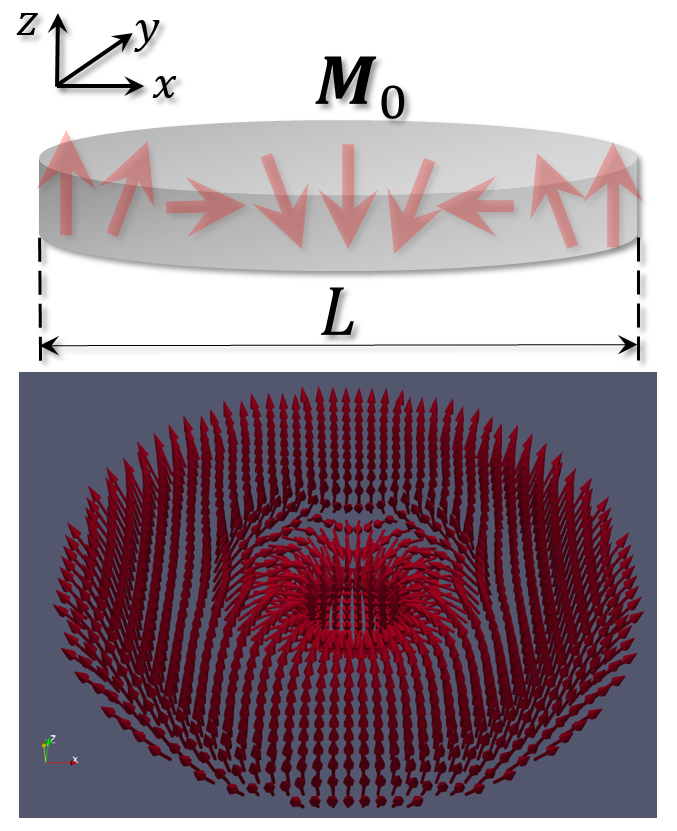}
	\caption{(Top) Thin-disk geometry, 
		and (bottom) classical micromagnetic Skyrmion ground state for disk with perpendicular anisotropy.}
	\label{fig:Disk_geometry_and_groud_state_Skyrmion}
\end{figure}

We treat here the quantum thermal fluctuations at the thermodynamic equilibrium around a noncollinear Bloch Skyrmion ground state in a thin disk of diameter $L=100$ nm and thickness $d=0.4$ nm with maximum magnetization $M_0=580$ kA/m, exchange constant $A=15$ pJ/m, perpendicular anisotropy constant $K_1=0.8\times 10^6$ J/m$^3$ and interfacial DM interaction constant $D=3.7$ mJ/m$^2$ (the parameters are taken from Ref.\cite{Wang_2018}). The choice of interfacial DM interaction corresponds to setting nonzero only the coefficients $D_x^{xz}=D,D_x^{zx}=-D,D_y^{yz}=D,D_y^{zy}=-D$ in the Hamiltonian \eqref{eq:magnetization_hamiltonian}. The disk is discretized into a collection of $2.5\times 2.5\times 0.4$ nm$^3$ square prism cells and is numerically simulated using the same methods described before. The geometry of the disk along with the computed Skyrmion ground state is reported in Fig.\ref{fig:Disk_geometry_and_groud_state_Skyrmion}. 
A subset of the first nine computed normal modes profiles is reported in Fig.\ref{fig:computed normal modes disk 100nm}(b). 

\begin{figure}[!t]
	\centering (a) \\
	\includegraphics[width=8.5cm]{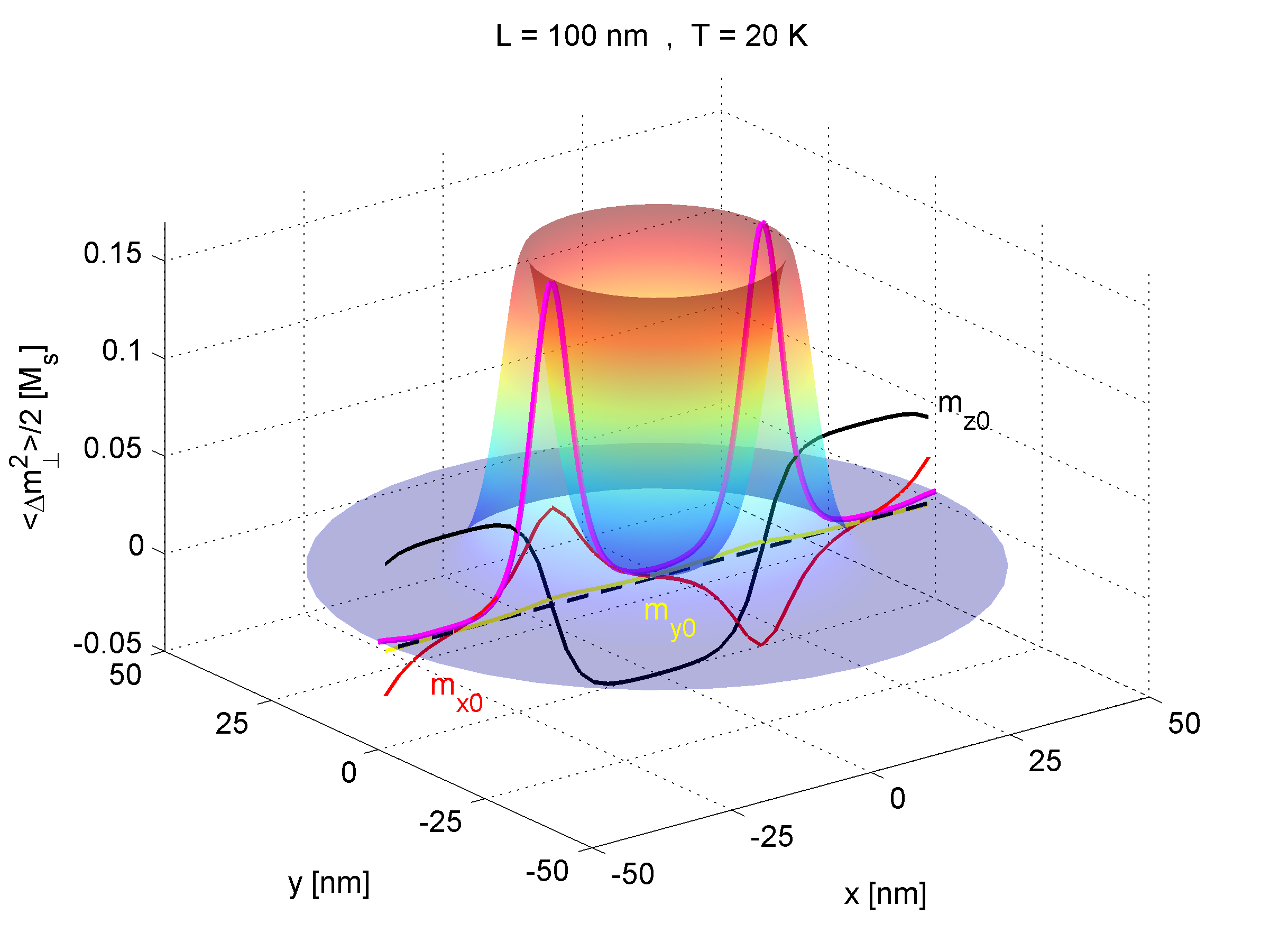} \\ (b) \\
	\includegraphics[width=8.5cm]{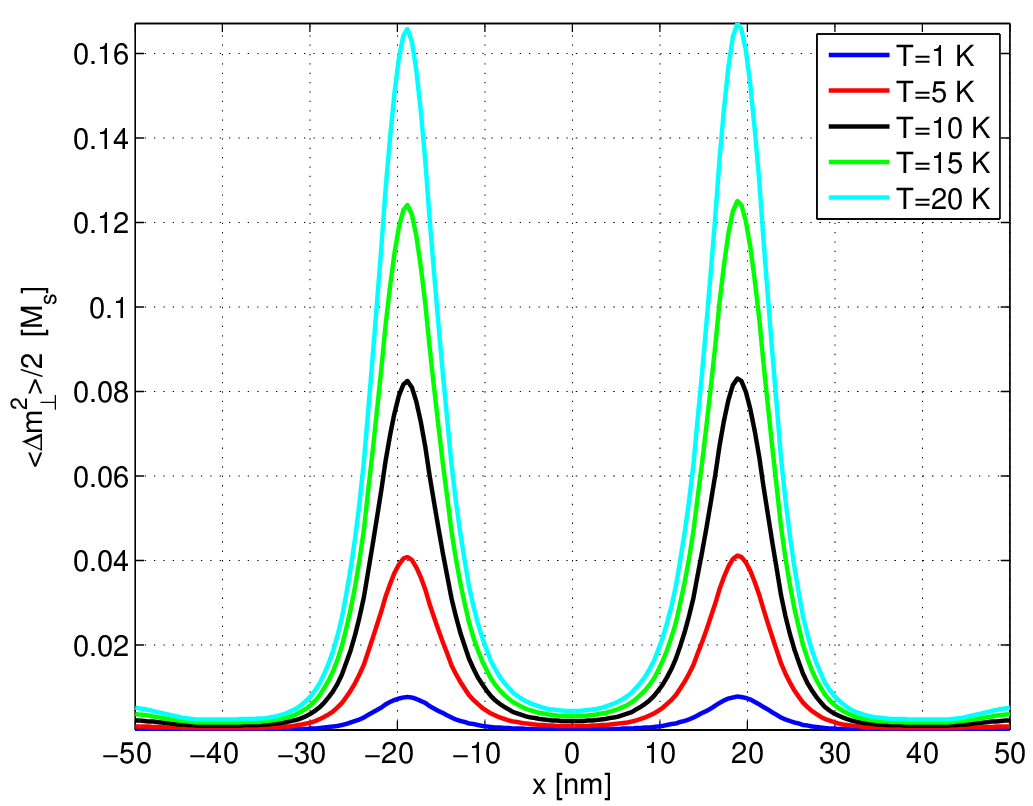}
	\caption{Profile of thermal average of $(\hat{M}_\bot^2(\xvec)/M_0^2)/2$ for a thin film with Skyrmion ground state. (a) the color map represents the transverse mean thermal deviation at $T=20$ K, magenta line refers to a line-cut along the $x$-axis. Solid black, yellow, red lines report line-cut along $x$ at $y=L/2=50$ nm of $m_{0}^z,m_{0}^y,m_{0}^x$ for the Bloch Skyrmion equilibrium profile (notice that they are scaled for visualization purposes). (b) Thermal deviation line-cut along $x$ as function of temperature.}
	\label{fig:Skyrmion_Disk_profile_mperp_vs_T_vs_L}
\end{figure}

The spatial profile of the thermal deviation $\hat{M}_\bot^2(\xvec)/2$ is reported in Fig.\ref{fig:Skyrmion_Disk_profile_mperp_vs_T_vs_L}. The Bloch Skyrmion ground state has rotational symmetry (see the line-cut diagram at $y=L/2=50$ nm of the equilibrium components $m_{0}^z(x),m_{0}^y(x),m_{0}^x(x)$ reported as black, yellow and red solid lines in Fig.\ref{fig:Skyrmion_Disk_profile_mperp_vs_T_vs_L}(a)). The thermal deviation along the same line-cut is depicted in solid magenta. One can clearly infer from the figure that, when the out-of-plane magnetization (black) changes from $+1$ (left Skyrmion edge $x<0$) to $-1$ (Skyrmion core), the $m_{0}^x$ component (red) grows towards +1.   Consequently, in that region the thermal deviation increases and is mainly oriented along the $y$ axis. Similar reasoning can be done for the other transition occurring for $x>0$.

In addition, we report in Fig.\ref{fig:Skyrmion_Disk_profile_mperp_vs_T_vs_L}(b) the line-cut diagram along $x$ of the thermal deviation for several temperature values $T=1,5,10,15,20$ K.

\begin{figure}[!t]
	\centering (a) \\
	\includegraphics[width=8.5cm]{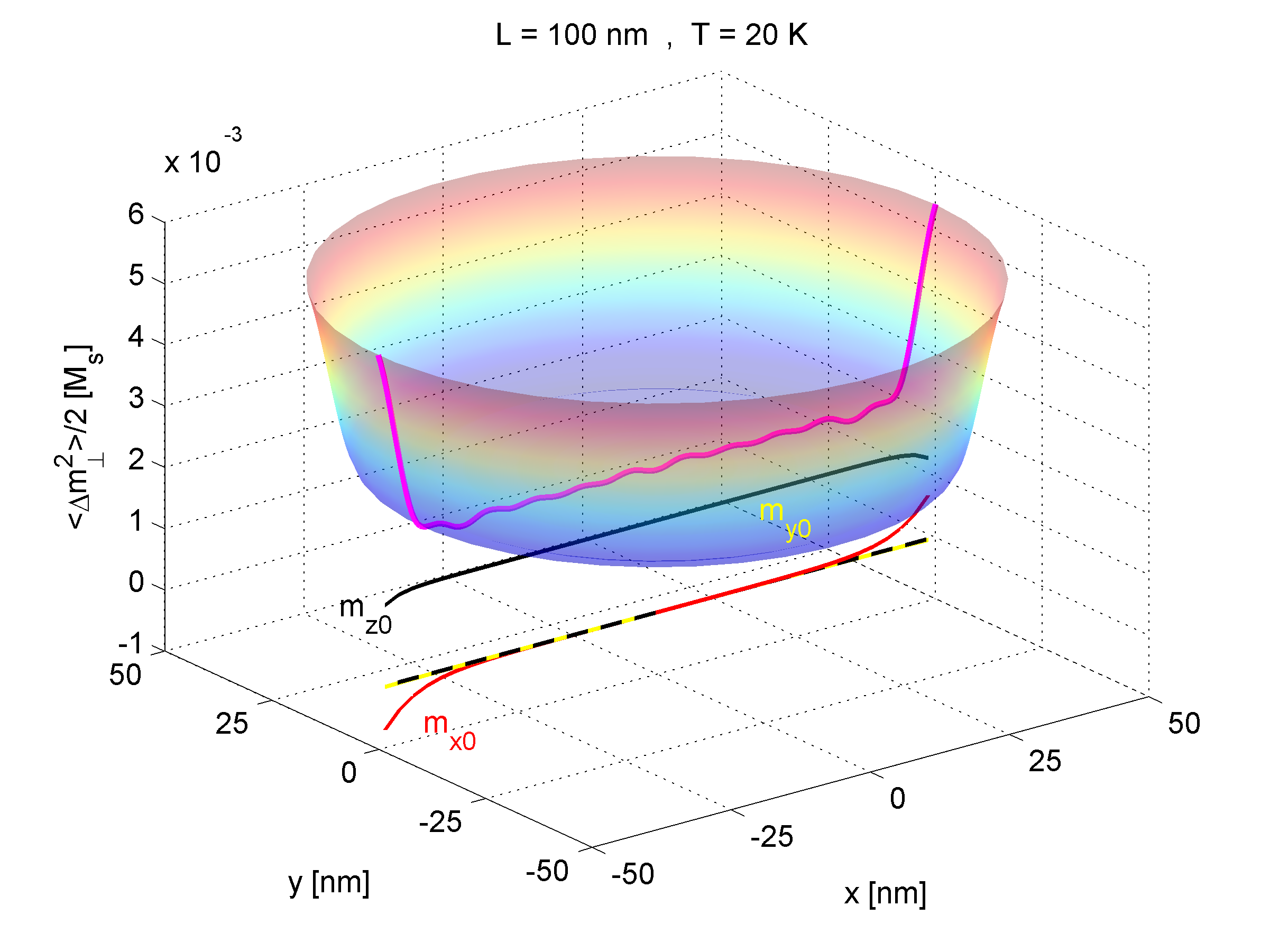} \\ (b) \\
	\includegraphics[width=8.5cm]{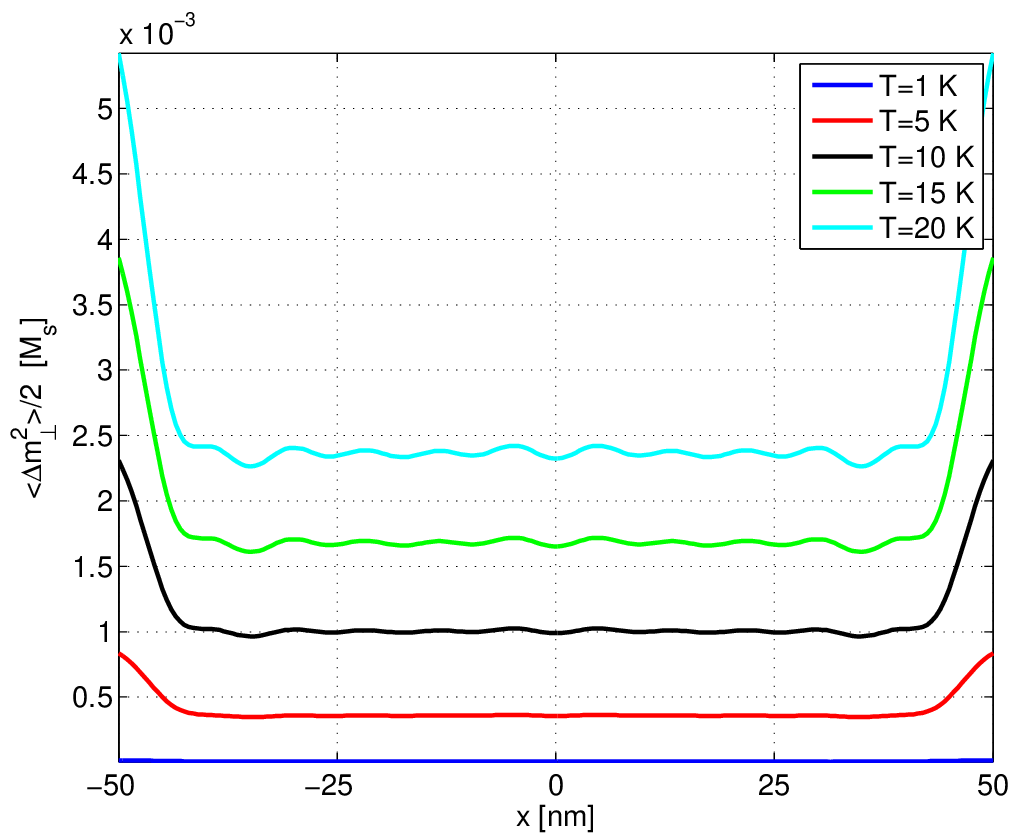}
	\caption{Profile of thermal average of $(\hat{M}_\bot^2(\xvec)/M_0^2)/2$ for a thin film with quasi-uniform ground state. (a) the color map represents the transverse mean thermal deviation at $T=20$ K, magenta line refers to a line-cut along the $x$-axis. Solid black, yellow, red lines report line-cut along $x$ at $y=L/2=50$ nm of $m_{z0},m_{y0},m_{x0}$ for the quasi-uniform equilibrium profile (notice that they are scaled for visualization purposes). (b) Thermal deviation line-cut along $x$ as function of temperature.}
	\label{fig:fm_100nm_Disk_profile_mperp_vs_T_vs_L}
\end{figure}

We also compare the above calculations with those related to a different ground state, hereafter referred to as quasi-uniform equilibrium state, for the thin disk with the same material parameters. The considered ground state has quasi-uniform magnetization mainly oriented along the positive $z$ axis (see the solid black, red and yellow lines in the panel (a) of fig.\ref{fig:fm_100nm_Disk_profile_mperp_vs_T_vs_L}) as a result of the competion between the perpendicular anisotropy and the DM interactions. A subset of the first nine computed normal modes profiles is reported in Fig.\ref{fig:computed normal modes disk 100nm}(c).

\begin{figure}[!t]
	\centering 
	\includegraphics[width=8cm]{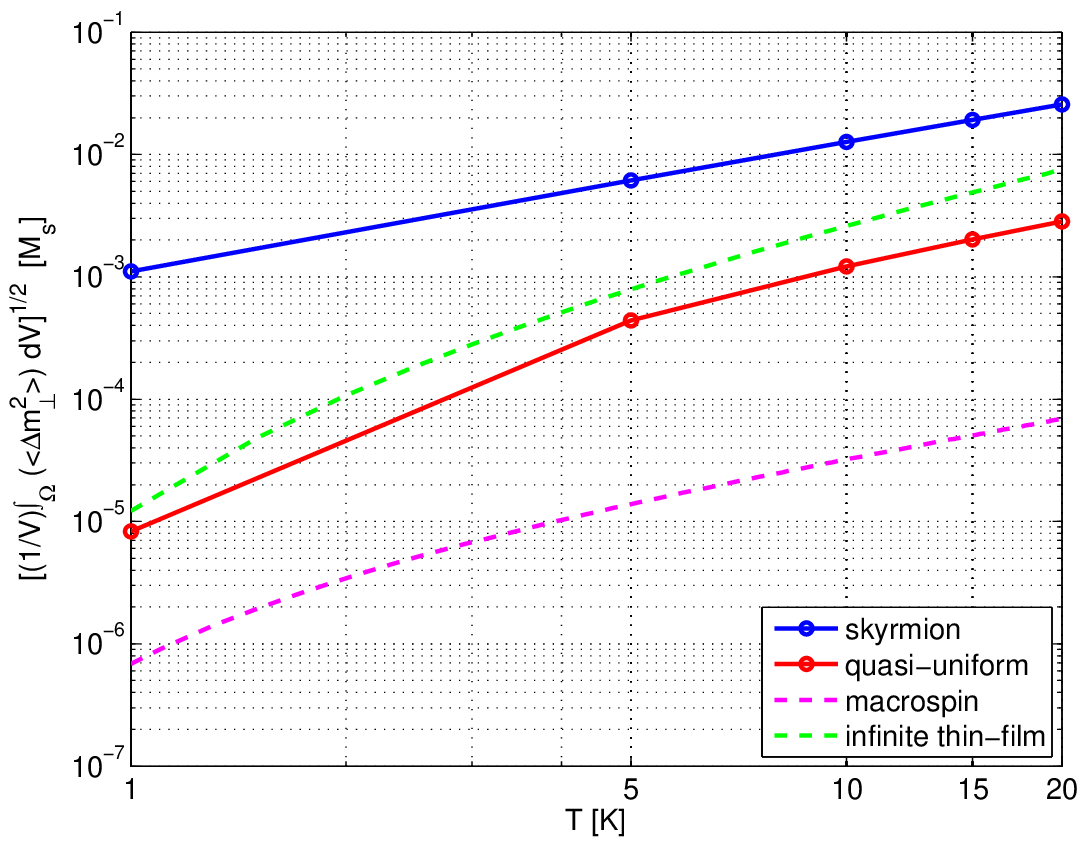} 
	\caption{Comparison of thermal mean square values  $(\hat{M}_\bot^2(\xvec)/M_0^2)/2$ for the thin disk with thickness 5 nm and diameter $L=100$ nm as function of temperature $T$. Solid blue line refers to the result for the skyrmion ground state, red line refers to the quasi-uniform equilibrium state, magenta line refers to the analytical calculation in the macrospin limit.}
	\label{fig:Cmp_Disk_thermal_mean_vs_T}
\end{figure}

Figures \ref{fig:fm_100nm_Disk_profile_mperp_vs_T_vs_L}(a)-(b) also report the spatial profile of the thermal deviation $(\hat{M}_\bot^2(\xvec)/M_0^2)/2$ as function of temperature that can be compared with that associated with the Skyrmion ground state (see Figs.\ref{fig:Skyrmion_Disk_profile_mperp_vs_T_vs_L}(a)-(b)). One can see that, for the quasi-uniform ground state, the thermal deviation is almost constant in the interior of the disk while it becomes more pronounced at the edges. In order to compare the overall thermal deviation for the two different ground states, we have reported the behavior of the mean square value of $(\hat{M}_\bot^2(\xvec)/M_0)/2$ over the disk in Fig.\ref{fig:Cmp_Disk_thermal_mean_vs_T} as function of temperature, which shows that the thermal deviation is significantly larger for the Skyrmion (blue line) than for the quasi-uniform (red line) ground state. A quick comparison of lower order normal modes spectra of the two cases in Figs.\ref{fig:computed normal modes disk 100nm}(b)-(c) suggests that the much larger thermal deviation occurring for the Skyrmion ground state, for which the lowest modes have much smaller frequencies/energies ($\sim 1$ GHz) than those arising in the quasi-uniform ground-state ($>57$ GHz), is owing to the much larger weights in the summation within formula \eqref{eq:quantum thermal mean square M_perp} involving the Bose statistics average occupation numbers at temperature $T$ (see eq.\eqref{eq:bose statistics average occupation number}). One can also see that the result for the quasi-uniform state is consistent with the analytical calculation performed in the macrospin and infinite thin-film limits (magenta and green dashed lines). 

\section{Conclusions}
	\label{sec:conclusions}
	
This work introduces a quantum-field-based micromagnetic approach for deriving the magnon spectrum 
in finite nanostructures. The derivation applies to a general quantum micromagnetic Hamiltonian 
that includes all relevant interactions, such as isotropic and anisotropic exchange, 
Dzyaloshinsky-Moriya (DM) interaction, dipolar interactions, anisotropy, and Zeeman effects. 
This formalism is well-suited for addressing spatially nonuniform ground states and accurately 
accounts for boundary effects. The computations presented in the paper demonstrate that edge effects induce 
quantitative changes in magnetization dynamics that cannot be captured by plane wave formalism alone.

The proposed approach extends the classical micromagnetic framework into the quantum regime. 
This advancement is expected to be of interest to researchers using classical micromagnetics to model 
systems and devices in fields like magnonics, spintronics, and related areas, particularly those 
interested in incorporating quantum effects into their models. This paper represents an initial step 
in this direction by addressing the case of non-interacting magnons.

The framework established here can be extended in multiple directions. Key examples include the quantum 
treatment of resonance under both transverse and longitudinal pumping, a perturbative analysis 
of magnon scattering with higher-order terms in the Hamiltonian expansion, and the study of relaxation 
processes due to interactions with a thermal reservoir. These topics will be explored in future studies, 
building on the theory of free magnon dynamics established here.

\section*{Acknowledgement}
The authors would like to express their gratitude to G. Bertotti, G. Miano, 
for their insightful discussions and valuable suggestions, which contributed 
to the clarification of the ideas presented in this paper. This work was partially 
supported from the Italian Ministry of University and Research, 
PRIN2020 funding program, grant number 2020PY8KTC, and partially supported
by the European Commission’s Horizon 2020 Framework Programme 
under Contract No. 899646 (k-Net).

\appendix

\section{Additional material on the derivation
of linearized quantum Landau-Lifshitz equation}
\label{appendix:expansion_of_H}

 \subsection{Expansion of the Hamiltonian around the ground state in the large $S$ limit}
\label{appendix:expansion_of_H_1}

The quantum Hamiltonian \eqref{eq:magnetization_hamiltonian}
can be written in the following form:
 \begin{align}
	\label{eq:H_in_terms_of_W}
	H[\hat{\Mvec}] = W[\hat{\Mvec},\hat{\Mvec}] - \int_\Omega \mu_0\hat{\Mvec}\cdot \Hveca dV \, ,
\end{align}
where $W[\,\cdot\,,\,\cdot\,]$ is a bilinear form associated to 
quadratic terms of the Hamiltonian \eqref{eq:magnetization_hamiltonian} and 
it is given by the following formula 
	\begin{align}
		\nonumber
&		{W}[\hat{\Fvec}, \hat{\Gvec}]  = \int_\Omega  dV  \left\{ \frac{A_{kh}^{uv}}{M_0^2}\, 
\frac{\partial \hat{F}^u}{\partial x_k}\frac{\partial \hat{G}^v}{\partial x_h} +\right.  	\\
\nonumber
&	\left.		\qquad\qquad\qquad	+  \frac{1}{2}{ \frac{D_{k}^{uv}}{M_0^2}   } \left( \! \hat{ F}^u \frac{\partial \hat{G}^v}{\partial x_k}
- \frac{\partial \hat{F}^u}{\partial x_k} \hat{ G}^v  \!\right)+ \right. \\ 
\label{eq:magnetization_hamiltonian_bilinear_quantum_0}
& \left.	\qquad\qquad\qquad \quad 	- \,\hat{F}^u \frac{K^{uv}}{M_0^2}\hat{G}^v  
		- \frac{\mu_0}{2}\hat{\Fvec }\cdot \Hvecm[\hat{\Gvec}]  \right\} \, ,
	\end{align}
where $\hat{\Fvec}(\xvec)$ and $\hat{\Gvec}(\xvec)$ are two generic quantum fields 
the components of which do not necessarily commute. 

The magnetostatic field operator present in the last term of 
eq.\eqref{eq:magnetization_hamiltonian_bilinear_quantum_0} is given by
the eqs.\eqref{eq:magnetostatic_field} , \eqref{eq:Green_fun_Dyadics_0}.
By using the fact that $N^{uv}(\xvec)$ in eq.\eqref{eq:Green_fun_Dyadics_0} is a symmetric tensor 
and it even with respect to $\xvec$, i.e.  $N^{uv}(-\xvec) =N^{uv}(\xvec)$, it can be proven that
\begin{align}
	\label{eq:Helmholtz_th_b_0}
	\int_\Omega  dV  \left\{	\hat{\Fvec}\cdot \Hvecm[\hat{\Gvec}]  \right\} 
	= \int_\Omega  dV  \left\{	\Hvecm[\hat{\Fvec}]\cdot \hat{\Gvec}  \right\} \, ,
\end{align}
and this is the magnetostatic reciprocity theorem for quantum fields.

Consider now the case when  $\hat{\Fvec}(\xvec)$ and $\hat{\Gvec}(\xvec)$  are
Hermitian operators, with possibly non-commuting Cartesian components. By using magnetostatic reciprocity 
\eqref{eq:Helmholtz_th_b_0}, the symmetry of the tensors $A^{uv}_{kh}$ and $K^{uv}$ and the anti-symmetry 
of the tensor $D_k^{uv}$ (with respect to indices $u,v$), we can easily check that the operator 
\eqref{eq:magnetization_hamiltonian_bilinear_quantum_0} satisfy the following condition 
$({W}[\hat{\Fvec}, \hat{\Gvec}] )^\dag = {W}[\hat{\Gvec}, \hat{\Fvec}]$. And this in turn implies 
that  $W[\hat{\Fvec},\hat{\Fvec}]$, $W[{\Vvec},\hat{\Fvec}]$ are Hermitian,  where  ${\Vvec}$ is a classical field.

The expansion \eqref{eq:expansion_of_H_M} can be expressed in terms of
bilinear form \eqref{eq:magnetization_hamiltonian_bilinear_quantum_0}.
To this end, we first rewrite expansion \eqref{eq:M03_expansion_over_GS} 
in the following form:
\begin{align}
	\label{eq:expansion_of_Mquantum}
	\hat{\Mvec}(\xvec) = \Mvec_0(\xvec) + \hat{\Mvec}_\bot(\xvec)+ \hat{\Vvec}_0(\xvec)	+
	\hat{\Wvec}_0(\xvec) \, ,
\end{align}
where
\begin{align}
	\label{eq:def_of_v0}
	&    	\hat{\Vvec}_0(\xvec) =  \Mvec_0(\xvec) \left[ - \frac{(\hat{\Mvec}_{\bot}(\xvec))^2 }{2M_0^2}\right] \, ,\\
	\label{eq:def_of_W0}
	& 	\hat{\Wvec}_0(\xvec) =	\Mvec_0(\xvec) {F}\left(\hat{\Mvec}_{\bot}(\xvec)/M_0\right)\, ,
\end{align}
$F(u)= \sqrt{1-u^2}-1+u^2/2$ is of order $\mathcal{O}(u^4)$, 
and where the Cartesian components of $\hat{\Mvec}_{\bot}(\xvec)$ satisfy the boundary conditions \eqref{eq:bc_m_perp}.
Notice that $\hat{\Vvec}_0(\xvec)$ is of second order in $\hat{\Mvec}_{\bot}(\xvec)/M_0$ while
$\hat{\Wvec}_0(\xvec)$ contains fourth and higher order powers of $\hat{\Mvec}_{\bot}(\xvec)/M_0$
In addition, $\hat{\Vvec}_0(\xvec)$  and $\hat{\Wvec}_0(\xvec)$ commute:
\begin{align}
	[\hat{\Vvec}_0(\xvec), \hat{\Wvec}_0(\xvec)]=0\, .
\end{align}
By substituting eq.\eqref{eq:expansion_of_Mquantum} in eq.\eqref{eq:H_in_terms_of_W},
we obtain that the first three terms of the expansion \eqref{eq:expansion_of_H_M} have the expressions
\begin{align}
	\label{eq:zeroth_oder_H_exp}
		H[\Mvec_0] = W[ \Mvec_0, \Mvec_0]  -  \int_\Omega \mu_0{\Mvec}_0\cdot \Hveca dV \, ,
\end{align}
\begin{align}
		\label{eq:first_oder_H_exp}
H_1[\hat{\Mvec}_\bot] = 2\, W[ \Mvec_0,\hat{\Mvec}_\bot] -  \int_\Omega \mu_0\hat{\Mvec}_\bot\cdot \Hveca dV \, , 
\end{align}
\begin{align}
		\nonumber
		&	 H_2[\Mvec_\bot] =W[ \hat{\Mvec}_\bot, \hat{\Mvec}_\bot]  + \\
			\label{eq:second_oder_H_exp}
		& \qquad \qquad \qquad +2 W[ \Mvec_0, \hat{\Vvec}_0] 	-  \int_\Omega \mu_0\hat{\Vvec}_0 \cdot \Hveca dV 	\,, 
\end{align}
where we used the fact that components of $\Mvec_0$ commute with the components of $\hat{\Mvec}_\bot$
and  $\hat{\Vvec}_0$. The zeroth order term $H[\Mvec_0]$ is  simply the micromagnetic energy 
corresponding to the classical ground state.

As far as the term $W[ \Mvec_0,\hat{\Mvec}_\bot]$ appearing in  $H_1[\hat{\Mvec}_\bot]$
is concerned, we can use the following result
\begin{align}
	\label{eq:quadratic_after_int_by_parts_0}
	W[\hat{\Mvec}_0,\hat{\Fvec}]  =  \int_\Omega  dV  \left\{ \frac{\mu_0}{2} \hat{\Fvec} \cdot\Dvec[{\Mvec}_0] \right\} \, ,
\end{align}
where the operator $\Dvec[\,\cdot\,]$ is defined by the following expression
\begin{widetext}
	\begin{align}
		\label{eq:Dvec_operator}
		-\Dvec[\hat{\Fvec}] =
		\evec_v \frac{1}{\mu_0}\left[ \frac{\partial}{\partial x_k}\left(
		\frac{2 A_{kh}^{uv}}{M_0^2} \frac{\partial \hat{F}^u}{\partial x_h} \right)  +
		\frac{\partial}{\partial x_k}\left( \frac{ D_{k}^{uv}}{M_0^2} \hat{F}^u \right)
		+ \frac{D_{k}^{vu}}{M_0^2}\frac{\partial \hat{F}^u }{\partial x_k } 
		+  \frac{{2}K^{uv}}{M_0^2}  {\hat{F}^u}\right] + \Hvecm[\hat{\Fvec}] \, ,
	\end{align}
\end{widetext}
and where $\hat{\Fvec}=\hat{\Fvec}(\xvec)$ is a generic quantum field.
The identity \eqref{eq:quadratic_after_int_by_parts_0} can be proven by using integration  by parts, 
boundary conditions \eqref{eq:bc_m_eq} for $\Mvec_0$, reciprocity theorem for magnetostatics (see eq.\eqref{eq:Helmholtz_th_b_0})
and the symmetry of the anisotropic tensor $K^{uv}$.

By  using \eqref{eq:quadratic_after_int_by_parts_0} into eq.\eqref{eq:first_oder_H_exp},
we arrive to the following expression
\begin{align}	
\label{eq:H1_final}
H_{1}[{\hat{\Mvec}_{\bot}} ] =  \int_{\Omega}  dV  \left\{ -\mu_0\Heff[\Mvec_0] \cdot \hat{\Mvec}_\bot  \right\}	\, ,
\end{align}
and since $\Heff[\Mvec_0]$ is aligned with $\Mvec_0$ (see eq.\eqref{eq:Brown_equilibrium})
$H_{1}[{\hat{\Mvec}_{\bot}} ]=0$. The quadratic term, given by eq.\eqref{eq:second_oder_H_exp}, 
can be written more explicitly as
\begin{align}
\nonumber
			&H_2[{\hat{\Mvec}_{\bot}} ] = \\
					\label{eq:H_2_of_expansion}
			& \qquad W[\hat{\Mvec}_\bot,\hat{\Mvec}_\bot] + 
\int_\Omega  dV  \left\{ \frac{\mu_0}{2} \lambda_0 (\hat{\Mvec}_\bot)^2 \right\} \, ,
\end{align}
where, we have used the identity
	\begin{align}	
	\nonumber
	&		    2W[ \Mvec_0, \hat{\Vvec}_0] 	-  \int_\Omega \mu_0\hat{\Vvec}_0 \cdot \Hveca dV 	 = \\
	\label{eq:H2_v0}
	& \qquad\qquad 		\int_{\Omega}  dV  \left\{ -{\mu_0 }\Heff[\Mvec_0] \cdot \hat{\Vvec}_0 \right\}	\, ,	
\end{align}
the fact that  $\Heff[\Mvec_0] = \lambda_0 \Mvec_0$ 
(see eq.\eqref{eq:Brown_equilibrium}) and the definition \eqref{eq:def_of_v0} of $\hat{\Vvec}_0$.
The derivation of equation \eqref{eq:H2_v0} is based on eq.\eqref{eq:quadratic_after_int_by_parts_0} 
used already to derive eq.\eqref{eq:H1_final}. The explicit expression of eq.\eqref{eq:H_2_of_expansion}
is reported in eq.\eqref{eq:Quadratic_Hamiltonian_M}.

The higher order terms in the expansion \eqref{eq:expansion_of_H_M} 
are expressed in terms of the bilinear form \eqref{eq:magnetization_hamiltonian_bilinear_quantum_0}
as follows
\begin{align}
	\label{eq:third_order_in_H_exp}
	&   H_3[\Mvec_\bot] = W[ \hat{\Mvec}_\bot, \hat{\Vvec}_0] + W[ \hat{\Vvec}_0, \hat{\Mvec}_\bot] \, ,\\
	\nonumber
	&   H_{\ge 4}[\Mvec_\bot] = 	-  \int_\Omega \mu_0\hat{\Wvec}_0 \cdot \Hveca dV +	\\
	\nonumber
	&\quad  W[ \hat{\Vvec}_0, \hat{\Vvec}_0] + 2\,W[{\Mvec}_0, \hat{\Wvec}_0]  +W[\hat{\Mvec}_\bot, \hat{\Wvec}_0]+\\
		\label{eq:fourth_order_in_H_exp}
	& \quad W[\hat{\Wvec}_0,\hat{\Mvec}_\bot] 
	 + 2 \, W[\hat{\Vvec}_0, \hat{\Wvec}_0] +   W[\hat{\Wvec}_0,\hat{\Wvec}_0] \, . 
\end{align}
In order to understand the scaling of these different terms as function of the parameter $M_0$, we introduce
the following dimensionless Hamiltonian operator
\begin{align}
	\label{eq:magnetization_hamiltonian_adim}
	h[\hat{\mvec}] =  \left( \frac{1}{2} \mu_0 M_0^2 |\Omega| \right)^{-1} H[\hat{\Mvec}] \, ,
\end{align}
where $|\Omega|$ is the volume of the region $\Omega$ and 
\begin{align}
	\hat{\mvec}(\xvec) = {\hat{\Mvec}(\xvec)}/{M_0} \, , \quad
	\hvec_\text{a}(\xvec)= \Hvec_\text{a}(\xvec)/{M_0}
\end{align}
are the normalized versions of the field $\hat{\Mvec}(\xvec)$, and $\Hveca(\xvec)$, respectively.
We can then expand the normalized Hamiltonian $h[\hat{\mvec}]$, as we did for $	H[{\hat{\Mvec}}(\cdot) ]$,
obtaining
\begin{align}\nonumber
	h[{\hat{\mvec}} ] =  	h[{{\mvec}_0} ]  
	+	h_1[{\hat{\mvec}}_{\bot}] +
	h_2[{\hat{\mvec}}_{\bot}] +   \\	
	\label{eq:expansion_of_h_M}
	+h_3[{\hat{\mvec}}_{\bot}] +
	h_{\ge 4}[{\hat{\mvec}}_{\bot}] \, ,
\end{align}
where $\hat{\mvec}_{\bot}= \hat{\Mvec}_{\bot}/M_0$.
Then it is easy to realize, from eqs.\eqref{eq:zeroth_oder_H_exp}-\eqref{eq:fourth_order_in_H_exp}
and the bilinear nature of $W[\,\cdot\, , \, \cdot\,]$, that $h_k[{\hat{\mvec}}_{\bot}]$ is infinitesimal of
order $k$ with respect to ${\hat{\mvec}}_{\bot}$. For this reason, since 
$h_1[{\hat{\mvec}}_{\bot}]=0$, the dominant nonconstant term for large values of $M_0$ is $h_2[{\hat{\mvec}}_{\bot}]$.

\subsection{Derivation of the torque form of linearized quantum Landau-Lifshitz equation}
\label{appendix:expansion_of_H_2}

In order to compute explicitly the right-hand-side of eq.\eqref{eq:Heisenberg_M_linearized},
it is useful to compute first the following commutator:
\begin{align}
	\nonumber
	&		[\hat{M}_\bot^s(\xvec),\hat{M}_\bot^u(\xvec^\prime) ]  = \\
	\label{eq:comm_rels_app_1}
	&             \qquad\qquad   = e_{0a}^{s}(\xvec) e_{0b}^{u}(\xvec^\prime)[\hat{M}^{0a}(\xvec),\hat{M}^{0b}(\xvec^\prime) ] \, .
\end{align}
By using eq.\eqref{eq:commu_rels_on_M_a_0_tris}, the last equation can be written as
\begin{align}
	[\hat{M}_\bot^s(\xvec),\hat{M}_\bot^u(\xvec^\prime) ] 
	= i({\gamma \hbar})M_0   \hat{\mathbbm{1}}  	\Lambda_0^{su}(\xvec)  \delta(\xvec-\xvec^\prime) \, ,
\end{align}
where
\begin{align}
	\label{eq:Lambda_0_a}
	\Lambda_0^{su}(\xvec) =  e_{02}^{s}(\xvec) e_{01}^{u}(\xvec)- e_{01}^{s}(\xvec) e_{02}^{u}(\xvec)  \, .
\end{align}
It can readily seen from eq.\eqref{eq:Lambda_0_a} that
\begin{align}
	\label{eq:Lambda_0_b}
	\Lambda_0^{su}(\xvec) m_0^u(\xvec) = 0 \, ,
\end{align}
and that 
\begin{align}
	\label{eq:Lambda_0_c}
	\evec_s	\Lambda_0^{su}(\xvec) e_{0a}^u(\xvec) = \epsilon_{ab3}e_{0b}^s(\xvec)  \, .
\end{align}
Equations \eqref{eq:Lambda_0_b}, \eqref{eq:Lambda_0_c} implies eq.\eqref{eq:Lambda_0_def}. 

A direct derivation of eq.\eqref{eq:LL_equation_spin-waves}
from eq.\eqref{eq:Heisenberg_M_linearized} can be obtained 
by using the expression ${H}_2[\hat{\Mvec}_\bot]$ given in eq.\eqref{eq:quadratic_Ham_D0_form}
that is based on the reciprocity property expressed in eq.\eqref{eq:reciprocity_D0_bot}.
Now, we want to compute the commutator at the right-hand-side of eq.\eqref{eq:Heisenberg_M_linearized}.  
We have that
\begin{align}
	\nonumber
	&	 \left[\hat{M}_\bot^s(\xvec) \, , \,   
	{H}_2[\hat{\Mvec}_\bot] \right] =\\
	\nonumber
	&  \int_\Omega dV^\prime  \frac{\mu_0}{2} \left\{  [ \hat{M}_\bot^s(\xvec),  \hat{M}^v_\bot(\xvec^\prime) ]
	\, D_{0\bot}^v[\hat{\Mvec}_\bot] + \right.\\
	\label{eq:reciprocity_proof_0}
	&\qquad   \left. \hat{M}^v_\bot(\xvec^\prime) D_{0\bot}^v\left[[\hat{M}^s(\xvec),\hat{M}^r_\bot(\xvec^\prime)]\evec_r\right] 
	\right\} \,,
\end{align}
where and $D_{0\bot}^v[\hat{\Mvec}]= \evec_v \cdot \Dvec_{0\bot}[\hat{\Mvec}]$.
By substituting the commutation relations for $\hat{\Mvec}_\bot(\xvec)$ (see eq.\eqref{eq:Mbot_comm_rel_cart}) in eq.\eqref{eq:reciprocity_proof_0}, 
and  using  eq.\eqref{eq:reciprocity_D0_bot},  we obtain  
\begin{align}
	\nonumber
	&	\int_\Omega dV^\prime  \frac{\mu_0}{2} \left\{ \left( i({\gamma \hbar})M_0   \hat{\mathbbm{1}}\right)  	\Lambda_0^{sv}(\xvec)  \delta(\xvec-\xvec^\prime)
	\, D_{0\bot}^v[\hat{\Mvec}_\bot] + \right. \\
	\nonumber
	&\qquad \qquad \left. D_{0\bot}^r[\hat{\Mvec}_\bot] \left( i({\gamma \hbar})M_0   \hat{\mathbbm{1}}  	\Lambda_0^{sr}(\xvec)  \delta(\xvec-\xvec^\prime) \right)
	\right\} = \\
	&  \qquad\qquad \qquad    =  (i\hbar\gamma \mu_0 M_0)\mvec_0(\xvec) \times \Dvec_{0\bot}[\hat{\Mvec}_\bot] \, .
\end{align}
This proves  equation \eqref{eq:LL_equation_spin-waves}.

\section{Quantum spin wave in infinite thin film using the plane wave formalism.}
\label{Appendix:Quantum spin wave treatment}

In this appendix we study the limiting case of the  problems treated in Sec.\,\ref{sec:applications} 
in which the ferromagnetic body is treated as an infinite ultra-thin film of thickness $d$.
We consider the two cases: in the first case the thin-film is saturated by
an in-plane applied field $H_a$ along the $x$-axis, and in the second case
the film is magnetized in the $z$-direction (perpendicular) to the film due to the
presence of perpendicular anisotropy. As we did in Sec.\,\ref{sec:applications},
we assume that exchange interaction is isotropic, and (when it is present) that
asisotropy is uniaxial with anisotropy asis perpedicular to the film plane. 
This means that $A^{uv}_{kh} = A \delta_{kh} \delta_{uv}$, 
$K=K_1 \evec_z\evec_z$. As in   Sec.\,\ref{sec:applications}, the magnetization field is 
assumed to not vary across the
thickness of the film. The derivation of the magnon spectrum  can be carried out analytically  
using plane waves formalism,based on the expansion \eqref{eq:plane_waves_diag_H_1},
with an in-plane wave vector:
$$\kvec = k_x \evec_x + k_y \evec_y\, .$$
The analytical results of plane wave formalism can be used as 
reference for the case of the disk with finite diameter.  

\subsection{Spin wave derivation in the case of in-plane magnetized thin film}

In this case, we assume that no DM neither anisotropy terms are present, this means that $D_k^{uv} =0$, $K_1=0$. 
In this conditions, the classical ground state is spatially uniform and given by $\Mvec_0= M_0 \evec_x$.
The applied field is spatially uniform and applied along the $x$-direction, i.e. $\Hveca = H_a \evec_x$.
The magnon spectrum is continuous and the associated frequencies can be expressed in terms of the dispersion relation
\begin{align}
	\omega(\kvec) = \gamma \mu_0 M_0 \sqrt{D_{0\bot}^{yy}(\kvec) D_{0\bot}^{zz}(\kvec)} \, ,
\end{align}
where 
 $D^{yy}_{0\bot}(\kvec), D^{zz}_{0\bot}(\kvec)$
are given by the formulas
\begin{align}
&	D^{yy}_{0\bot}(\kvec) = \frac{2A}{\mu_0 M_0^2}k^2 + \lambda_0 + [1 - S(kd)]\frac{k_y^2}{k^2}\, ,\\
&   D^{zz}_{0\bot}(\kvec) = \frac{2A}{\mu_0 M_0^2}k^2 + \lambda_0 + S(kd)\, , 
\end{align}
and where
\begin{align}
\label{eq:def_S_k_lambda_0}
	 k = \sqrt{k_x^2+k_y^2} \,,   \,\, S(kd) = \frac{1- e^{-kd}}{kd} \, , \,\, \lambda_0 = \frac{H_a}{M_0} .
\end{align}
Next we compute the, as in Sec.\,\ref{sec:applications}, the thermal equilibrium mean square values
of the out-of-plane ($z$-direction) and the in-plane ($y$-direction) components of $\hat{\Mvec}(x,y)$.
These are given by the following formulas:
\begin{align}
	\nonumber
	&	\langle (\hat{M}_\bot^z(x,y))^2 \rangle_T =  \\ 
\label{eq:inf_thin_fil_mq_mz}
	& \quad \frac{\mu_0 \hbar (\gamma  M_0)^2}{(2\pi)^2d} 
	\int_{\mathbb{R}^2} dk_x dk_y \left\{  \langle \hat{n}(\kvec) \rangle_T  \frac{D^{yy}_{0\bot}(\kvec)}{\omega(\kvec)} \right\} \, ,
\end{align}
and
\begin{align}
	\nonumber
&	\langle (\hat{M}^y_\bot(x,y))^2 \rangle_T =  \\ 
\label{eq:inf_thin_fil_mq_my}
& \quad \frac{\mu_0 \hbar (\gamma  M_0)^2}{(2\pi)^2d} 
\int_{\mathbb{R}^2} dk_x dk_y \left\{ \langle \hat{n}(\kvec) \rangle_T   \frac{D^{zz}_{0\bot}(\kvec)}{\omega(\kvec)} \right\} \, ,
\end{align}
where
\begin{align}
		\langle \hat{n}(\kvec) \rangle_T  = \frac{1}{e^{\beta\hbar\omega(\kvec)}-1} \, .
\end{align}
Since in this case the physical system is invariant with respect to in-plane spatial translations,
the above results do not depend on the position.
The values obtained by the expressions \eqref{eq:inf_thin_fil_mq_mz}, \eqref{eq:inf_thin_fil_mq_my}
are compared  in Sec.\,\ref{sec:applications} with the analogous quantities computed for finite
diameter thin disk.

\subsection{Spin wave derivation in the case of out-of-plane magnetized film}

Here, for comparison with the second set of numerical computations
presented in Sec.\,\ref{sec:applications},  we report the formulas for the case of infinite 
thin film that admits a spatially uniform ground state in the $z$-direction, $\Mvec_0 = M_0 \evec_z$ 
caused by a perpendicular anisotropy that is defined by a tensor $K^{uv}$ in  which the only nonzero 
entry is $K^{zz} = K_1$. The formulas that are reported below  are not affected by the 
presence of the DM interaction as far as the nonzero entries in the tensor $D_k^{uv}$ are 
only $D_x^{xz} = - D_x^{xz}$ and $D_y^{yz}=-D_y^{zy}$ as it is assumed
in Sec.\,\ref{sec:applications}. In the present case,
the continuous  magnon spectrum is characterized by the dispersion relation
\begin{align}
\label{eq:dispersion_rel_perp}
	\omega(\kvec) = \gamma \mu_0 M_0 \sqrt{D_{0\bot}^{xx}(\kvec) D_{0\bot}^{yy}(\kvec)- (D_{0\bot}^{xy}(\kvec))^2 } \, ,
\end{align} 
where
\begin{align}
	\label{eq:D0botkvec_xx}
	&	D^{uv}_{0\bot}(\kvec) = 
    \left( \frac{2A k^2}{\mu_0 M_0^2} + \lambda_0\right) \delta_{uv} + 
    [1 - S(kd)]\frac{k_u k_v}{k^2}\, ,
\end{align}
where $u,v=x,y$ ($D^{uv}_{0\bot}(\kvec)$ is $2\times2$ matrix), $k$ and  $S(kd)$ are as defined 
eq.\eqref{eq:def_S_k_lambda_0}, and where
\begin{align}
    \lambda_0 = {2K_1}/({\mu_0 M_0^2}) - 1\,.
\end{align}
It can readily realized, substituting formula 
\eqref{eq:D0botkvec_xx} in eq.\eqref{eq:dispersion_rel_perp}, that
$\omega(\kvec)$ is only a function of the magnitude $k$ of $\kvec$.
Next, we compute the the thermal equilibrium mean square values
of $x$ and $y$ components of $\hat{\Mvec}(x,y)$. Due to the rotational
symmetry of the problem, this two averages are the same and are given by the following formula:
\begin{align}
	\nonumber
	&	\langle (\hat{M}_\bot^x(x,y))^2 \rangle_T = \langle (\hat{M}^y_\bot(x,y))^2 \rangle_T = \\ 
	\label{eq:inf_thin_fil_mq_mz_perp}
	& \frac{\mu_0 \hbar (\gamma  M_0)^2}{2(2\pi)^2d} 
	\int_{\mathbb{R}^2} dk_x dk_y \left\{ \langle \hat{n}(\kvec) \rangle_T  
    \frac{\text{tr}(D^{uv}_{0\bot}(\kvec))}{\omega(\kvec)} \right\} \, ,
\end{align}
where $\text{tr}(\,\cdot\,)$ denotes the trace of a tensor.
Also in this case the invariance  of the system with respect to in-plane spatial translations,
leads to results that are not function of the position.
The values obtained by the expression \eqref{eq:inf_thin_fil_mq_mz_perp}
are compared  in Sec.\,\ref{sec:applications} with the analogous quantities 
numerically computed  for a thin disk of diameter $L$ and with 
out-of-plane ground state.


\newpage
	\begin{thebibliography}{99}
		
		\bibitem{Bloch_1930} F. Bloch, Zur Theorie des Ferromagnetismus, Z. Physik {\bf 61}, 206 (1930); 
		\bibitem{Bloch_1932} F. Bloch, Zur Theorie des Austauschproblems und der Remanenzerscheinung der Ferromagnetika, Z. Physik {\bf 74}, 295 (1932).
		\bibitem{Holstein_Primakoff_1940} T. Holstein, H. Primakoff, Field Dependence of the Intrinsic Domain Magnetization of a Ferromagnet, 
		Phys. Rev. {\bf 58}, 1098 (1940)
		\bibitem{Herring_Kittel_1951} C. Herring and C. Kittel, On the Theory of Spin Waves in Ferromagnetic Media
		Phys. Rev. {\bf 81}, 869 (1951); Erratum Phys. Rev. {\bf 88}, 1435 (1952)
		\bibitem{Dyson_1956} F.J. Dyson, General theory of Spin-Wave interactions, Phys. Rev. {\bf 102}, 1217 (1956)
		\bibitem{Kittel_book_1963} C. Kittel, Quantum theory of solids, Wiley,  (1963). 
	\bibitem{Akhiezer_book} A. I. Akhiezer,  V.G.  Bariakhtar,
       S.V. Peletminskii, Spin Waves, Interscience Publishers, New York (1968)
        
        \bibitem{Stancil_Prabhakar_book} D. D. Stancil, A. Prabhakar,
		Spin Waves: Theory and Applications, Springer  (2009).
		\bibitem{Sparks_book} M. Sparks, Ferromagnetic Relaxation Theory, McGraw-Hill (1964)., 
		
	
		
		\bibitem{Urazhdin_2014}
		S. Urazhdin, V. E. Demidov, H. Ulrichs, T. Kendziorczyk, T. Kuhn, J. Leuthold, G. Wilde, S. O. Demokritov 
		Nanomagnonic devices based on the spin-transfer torque, Nature Nanotechnology 9, p. 509 (2014)
		
		\bibitem{Demidov_2020}
		V. E. Demidov, S. Urazhdin, A. Anane, V. Cros, S. O. Demokritov,
		Spin–orbit-torque magnonics, J. Appl. Phys. {\bf 127}, 170901 (2020)
		
		\bibitem{Lenk_2011}
		B. Lenk, H. Ulrichs, F. Garbs, M. Münzenberg, 
		The building blocks of magnonics, Physics Reports,
		Volume {\bf 507}, Issues 4–5, 2011
		
		\bibitem{Barman_2021}
		A. Barman et al, The 2021 Roadmap on Magnonics 
		J. Phys.: Condens. Matter,{\bf 33}, 413001 (2021)
		
		\bibitem{Chumak_2015}
		A. Chumak,  V. Vasyuchka, V. Serga et al. 
		Magnon spintronics. Nature Phys {\bf 11}1, 453–461 (2015).
		
		\bibitem{Chernov_2020}
		A. I. Chernov, M. A. Kozhaev, D. O. Ignatyeva, E. N. Beginin, A. V. Sadovnikov, A. A. Voronov, D. Karki, M. Levy, V. I. Belotelov, 
		All-Dielectric Nanophotonics Enables Tunable Excitation of the Exchange Spin Waves,
		Nano Lett. 2020, 20, 7, 5259–5266, 2020
		
		\bibitem{Maksymov_2016}
		Ivan S. Maksymov,
		Magneto-plasmonic nanoantennas: Basics and applications,
		Reviews in Physics,
		Volume 1, 2016
		
		\bibitem{Yuan2022} H.Y. Yuan, Y. Cao, A. Kamra, R. A. Duine, P. Yan, 
		Quantum magnonics: When magnon spintronics meets
		quantum information science,  Physics Reports {\bf 965} 1 (2022).

		\bibitem{Dzyaloshinskii1958}
	    I. E. Dzyaloshinskii, A thermodynamic theory of “weak” ferromagnetism of antiferromagnetics,
		Journal of Physics and Chemistry of Solids {\bf 4}, 4 (1958)
		
		
		\bibitem{Moriya1960}
		T. Moriya, Anisotropic Superexchange Interaction and Weak
		Ferromagnetism, Phys. Rev. {\bf 120}, 91 (1960).
		
		\bibitem{Bogdanov1994}
		A. Bogdanov, A. Hubert,
		Thermodynamically stable magnetic vortex states
		in magnetic crystals, J. of Magn. and Magn. Mat. {\bf 138}  255-269 (1994)
		
		\bibitem{Fert2017} A. Fert, N. Reyren, V. Cros, 
		Magnetic skyrmions: advances in physics and potential applications. 
		Nat Rev Mater {\bf 2}, 17031 (2017). 
		
		\bibitem{Brown_1963}
		William Fuller Brown, Micromagnetics, Interscience Publishers (1963).
		
		
		\bibitem{Walker_1957} L. R. Walker, Magnetostatic Modes in Ferromagnetic Resonance,
		Phys. Rev. {\bf 105}, 390 (1957).
		
		\bibitem{Aharoni_1963} A. Aharoni, Exchange resonance modes in a ferromagnetic sphere,
		J. Appl. Phys. {\bf 69}  7762 (1991).
		
	
		
		
		\bibitem{Arias_2001} R. Arias, D.L. Mills, Theory of spin excitations and the microwave response of cylindrical ferromagnetic nanowires,
		 Phys. Rev. B {\bf 43}  134439 (2001); 
		 Phys. Rev. B {\bf 66}  149903(E) (2002).
		
		\bibitem{Arias_2004} R. Arias, D.L. Mills,  Magnetostatic modes in ferromagnetic nanowires, Phys. Rev. B {\bf 70} 094414 (2004).
		
		\bibitem{Arias_2005a} R. Arias, P. Chu, D.L. Mills, Dipole exchange spin waves and microwave response of ferromagnetic spheres, Phys. Rev. B {\bf 71} 224410 (2005).
		
		\bibitem{Arias_2005b} R. Arias, D.L. Mills,  Magnetostatic modes in ferromagnetic nanowires. II. A method for cross sections with very
		large aspect ratio, Phys. Rev. B {\bf 72},  104418 (2005).
		
		
		
		\bibitem{Labbe_1999} S. Labb\'{e}, P.-Y. Bertin, Microwave polarizability of ferrite particles
		with non-uniform magnetization, J. Magn. Magn. Mater. {\bf 206}, 93 (1999) .
		
		\bibitem{Vukadinovic_200} N. Vukadinovic, O. Vacus, M. Labrune, O. Acher, D. Pain,
		Magnetic Excitations in a Weak-Stripe-Domain Structure: A 2D Dynamic
		Micromagnetic Approach, Phys. Rev. Lett. {\bf 85} 2817 (2000).
		
		\bibitem{Grimsditch_2004a} M. Grimsditch, G.K. Leaf, H.G. Kaper, and D.A. Karpeev,
		R.E. Camley, Normal modes of spin excitations in magnetic nanoparticles, Phys. Rev. B {\bf69}, 174428 (2004).
		
		\bibitem{Grimsditch_2004b} M. Grimsditch, L. Giovannini, F. Montoncello, F. Nizzoli, G.K. Leaf and H.G. Kaper,
		Magnetic normal modes in ferromagnetic nanoparticles: A dynamical matrix approach, Phys. Rev. B {\bf 70}, 054409 (2004).
		
		\bibitem{Rivkin_2004} K. Rivkin, A. Heifetz, P.R. Sievert, and J.B. Ketterson, Resonant modes of dipole-coupled lattices, 
		Phys. Rev. B {\bf 70}, 184410 (2004).
		
		\bibitem{dAquino_2009} M. d’Aquino, C. Serpico, G. Miano, C. Forestiere,
		A novel formulation for the numerical computation of magnetization
		modes in complex micromagnetic systems, 
		Journal of Computational Physics {\bf 228}, p. 6130 (2009) 
		
		\bibitem{McMicheal_2005}
		R. D. McMichael, and M. D. Stiles, Magnetic normal modes of nanoelements, J. Appl. Phys. {\bf 97}, 10J901 (2005).
		
		\bibitem{Mills_2006} D.L. Mills, Quantum theory of spin waves in finite samples,
		JMMM {\bf 306} 16 (2006).
		
		\bibitem{Mills_2007} D.L. Mills, Spin Waves: History and a Summary of Recent
		Developments, Handbook of Magnetism and Advanced Magnetic Materials, S. Parkin, H.Kronmueller (eds),
		Wiley (2007) 
		
		\bibitem{Rastelli_2013} E. Rastelli, Statistical Mechanics of
		Magnetic Excitations, From Spin Waves to Stripes and Checkerboards,
		World Scientific, Singapore (2013)
		
	       \bibitem{Diep_2022} H.T. Diep,  
	    Quantum Spin-Wave Theory for Non-Collinear Spin Structures, a Review. Symmetry {\bf 14}, 1716 (2022)
		
		\bibitem{Schuetz_2003} F. Schütz, M. Kollar, and P. Kopietz,
		Persistent Spin Currents in Mesoscopic Heisenberg Rings,
		Phys. Rev. Lett. 91, 017205 (2003)
		
		\bibitem{Schuetz_2004a}
		F. Schütz, P. Kopietz, M.  Kollar, 
		What are spin currents in Heisenberg magnets ?, 
		Eur. Phys. J. B {\bf 41}, 557 (2004)
		
		\bibitem{Schuetz_2004b}
		F. Schütz, M. Kollar, and P. Kopietz,
		Persistent spin currents in mesoscopic Haldane-gap spin rings
		Phys. Rev. B {\bf 69}, 035313 (2004)
		

		
	
		
		\bibitem{Kreisel_2009}
		A. Kreisel, F. Sauli, L. Bartosch, and P. Kopietz, 
		Eur. Phys. J. B 71, 59–68 (2009)
		
		\bibitem{Spremo_2005} I. Spremo, F. Schuetz, P. Kopietz, V. Pashchenko,
		Magnetic properties of a metal-organic antiferromagnet on a distorted honeycomb lattice
		Phys. Rev. B {\bf72}, 174429 (2005)
		
			\bibitem{Kreisel_2008}
		A. Kreisel, F. Sauli, N. Hasselmann, and P. Kopietz,
		Quantum Heisenberg antiferromagnets in a uniform magnetic field: Nonanalytic magnetic field
		dependence of the magnon spectrum,	Phys. Rev. B {\bf 78}, 035127 (2008)
		
			\bibitem{Roldan-Molina_2014} A. Rold\`{a}n-Molina, M. J. Santander, \'A.\,S.\,N\'{u}n\~{e}z,
		J. Fern\'andez-Rossier, Quantum theory of spin waves in finite chiral spin chains,
		Phys. Rev. B {\bf 89}, 054403 (2014)
		
			\bibitem{Roldan-Molina_2015} A. Rold\`{a}n-Molina, M. J. Santander, \'A.\,S.\,N\'{u}n\~{e}z, J. Fern\'andez-Rossier, 
		Quantum fluctuations stabilize skyrmion textures, Phys. Rev. B {\bf 92}, 245436 (2015)
		
		\bibitem{Rueckriegel_2017} A. Rueckriegel, P. Kopietz,
		Spin currents, spin torques, and the concept of spin superfluidity
		Phys. Rev. B {\bf 95}, 104436 (2017)
		
	    \bibitem{Haraldsen_2009}
	    J.T. Haraldsen, R. S. Fishman,
		Spin rotation technique for non-collinear magnetic systems: application to the generalized Villain model,
		J. Phys.: Condens. Matter {\bf 21},  216001 (2009)
		
		\bibitem{Aharoni_Ferromagnetism}
		A. Aharoni, Introduction to the Theory of Ferromagnetism,
		Clarendon Press (2000)
  
  
        	\bibitem{Abert_2019}
        C. Abert, Micromagnetics and spintronics: models and numerical methods,		
        Eur. Phys. J. B {\bf 92}, 120 (2019) 
        
     
      \bibitem{Yuan2022a} H. Y. Yuan, W. P. Sterk, Akashdeep Kamra, and R.A. Duine,
      Master equation approach to magnon relaxation and dephasing, 
      Physical Review B {\bf 106}, 224422(2022)
      
     
        
        \bibitem{MaGICo} M. d'Aquino, Magnetization Geometrical Integration Code (MaGICo),\url{http://wpage.unina.it/mdaquino/index\_file/MaGICo.html}
        
          \bibitem{Heitler_1936} W. Heitler, The Quantum Theory of Radiation, Oxford (1935)
        

        \bibitem{Wang_2018} Wang, X.S., Yuan, H.Y., Wang, X.R. A theory on skyrmion size. Commun Phys 1, 31 (2018).
	\end{thebibliography}
\end{document}